\documentclass[manuscript,screen,timestamp,nonacm]{acmart}

\renewcommand{\subsubsectionautorefname}

\usepackage[yyyymmdd]{datetime}

\usepackage{appendix} 

\usepackage{array}
\usepackage{pdflscape}
\usepackage{tabularx}
\usepackage{booktabs}
\usepackage{array}
\usepackage{enumitem}
\usepackage{ltablex}

\newlist{tightitemize}{itemize}{1}
\setlist[tightitemize,1]{label=--, left=1em, labelsep=0.5em, itemsep=0pt, topsep=2pt, parsep=0pt, partopsep=0pt, align=left}

\usepackage{mdframed}
\definecolor{Gray}{gray}{0.95}
\mdfdefinestyle{contributionmdf}{skipabove=20pt, skipbelow=10pt, linewidth=0pt, innertopmargin=10pt, innerbottommargin=10pt, backgroundcolor=Gray}

\newenvironment{summarize}[1]{
	\begin{mdframed}[style=contributionmdf]
	#1
	\end{mdframed}
}

\usepackage{tabularx}
\usepackage{multirow}
\usepackage{dcolumn} 
\newcolumntype{d}[1]{D{.}{.}{#1}}

\usepackage{subcaption}
\newcommand{\quotes}[1]{\textit{``#1''}}

\usepackage{booktabs} 

\usepackage{todonotes}
\let\xtodo\todo
\renewcommand{\todo}[1]{\xtodo[inline,color=green!50]{#1}}

\AtBeginDocument{%
  }

\setcopyright{acmcopyright}
\copyrightyear{2018}
\acmYear{2018}
\acmDOI{XXXXXXX.XXXXXXX}

\acmConference[Conference acronym 'XX]{Make sure to enter the correct
  conference title from your rights confirmation emai}{June 03--05,
  2018}{Woodstock, NY}
\acmISBN{978-1-4503-XXXX-X/18/06}

\begin{document}

\title[An Empirical Large-Scale Study on LLM Use in Everyday Learning]{Conversational AI as a Catalyst for Informal Learning: An Empirical Large-Scale Study on LLM Use in Everyday Learning}
\settopmatter{authorsperrow=3}

\author{Nađa Terzimehić}
\orcid{0000-0001-6630-3512}
\affiliation{%
  \institution{Technical University of Munich}
  \city{Munich}
  \postcode{80335}
  \country{Germany}}
\email{nadja.terzimehic@tum.de}

\author{Babette Bühler}
\orcid{0000-0003-1679-4979}
\affiliation{%
  \institution{Technical University of Munich}
  \city{Munich}
  \country{Germany}
}
\email{babette.buehler@tum.de}

\author{Enkelejda Kasneci}
\orcid{0000-0003-3146-4484}
\affiliation{%
  \institution{Technical University of Munich}
  \city{Munich}
  \country{Germany}
}
\email{enkelejda.kasneci@tum.de}

\renewcommand{\shortauthors}{Terzimehić et al.}

\begin{abstract}
Large language models have not only captivated the public imagination but have also sparked a profound rethinking of how we learn. In the third year following the breakthrough launch of ChatGPT, everyday informal learning has been transformed as diverse user groups explore these novel tools. Who is embracing LLMs for self-directed learning, and who remains hesitant? What are their reasons for adoption or avoidance? What learning patterns emerge with this novel technological landscape? We present an in-depth analysis from a large-scale survey of 776 participants, showcasing that 88\% of our respondents already incorporate LLMs into their everyday learning routines for a wide variety of (learning) tasks. Young adults are at the forefront of adopting LLMs, primarily to enhance their learning experiences independently of time and space. Four types of learners emerge across learning contexts, depending on the tasks they perform with LLMs and the devices they use to access them. Interestingly, our respondents exhibit paradoxical behaviours regarding their trust in LLMs' accuracy and privacy protection measures. Our implications emphasize the importance of including different media types for learning, enabling
collaborative learning, providing sources and meeting the needs of different types of learners and learning by design.
\end{abstract}




\maketitle

\section{Introduction}

The mass adoption of large language models (LLMs), accelerated by the release of OpenAI's ChatGPT in November '22, has captivated both the public and the research community ever since. ChatGPT and similar LLMs, such as Gemini, Grok, DeepSeek, are capable of engaging in natural language dialogue and providing detailed responses across diverse topics. Numerous thought pieces and research studies have addressed the impact of these novel tools in several areas of our everyday lives, with remarkably education and learning being at the forefront of change~\cite{kasneci2023chatgpt, grassini2023shaping, pang2025llmification}. 

Prior work in Human-Computer Interaction (HCI) and educational sciences highlights both the promises and challenges of integrating AI into learning and education practices~\cite{kasneci2023chatgpt}, with some researchers naming it a \quotes{double-edged sword} for education~\cite{HADIMOGAVI2024chatgptBlessing}. On the one \textit{edge}, LLM-driven assistants have demonstrated the potential for enhancing learning engagement and effectiveness~\cite{CHIU2023systematic, lyu2024evaluating}, as well as facilitating personalized learning~\cite{weijers-etal-2024-quantifying, park2024empowering, CHIU2023systematic}. On the other edge, concerns persist regarding over-reliance \cite{STADLER2024cognitive, DARVISHI2024impact}, inaccuracies \cite{wang2024factuality}, biases \cite{Baker2022bias, lin2025implicitbiasllmssurvey}, or ethical issues~\cite{kasneci2023chatgpt,HADIMOGAVI2024chatgptBlessing}.

In light of LLMs' recently emerging capabilities for reasoning and reflection~\cite{berti2025emergent}, the impact of LLMs on learning bears the potential to be far more groundbreaking than initially anticipated: learners now have a tailored, around-the-clock learning assistant by their side~\cite{clarizia2018chatbots, kasthuri2021lifestyle, sessler2023peer}, to assist them in a plethora of \textit{cognitive} tasks. This presents an opportunity to democratize learning and embed it into everyday life scenarios -- i.e., everyday learning -- be it informal learning as day-to-day, often incidental learning; voluntary use in structured educational environments; or self-directed knowledge acquisition outside of formal curricula~\cite{Rogoff2016informal, Callanan2011Informal, schugurensky2000forms} for both professional and personal growth. 

Whereas the role of AI and, in particular, LLMs has been researched for formal education settings, informal learning settings have received less attention, especially at scale. We believe this to be an opportune time to explore the \textit{actual use} of LLMs in informal contexts now that the initial hype and excitement around LLMs have settled. 
Prior research works have explored the use and perceptions of LLMs in different learning settings, including a qualitative large-scale study with early ChatGPT adopters across different disciplines and learning areas~\cite{HADIMOGAVI2024chatgptBlessing}, a large-scale survey of researchers using LLMs as research tools~\cite{liao2024llmsresearchtoolslarge}, investigations into the integration and use of LLMs within formal education by students~\cite{sublime2024chatgpt} or examinations into the usability of LLMs~\cite{skjuve2023uxchatgpt}. While these studies provide notable findings, they predominantly focus on either qualitative insights or specialized user groups, leaving a gap in understanding the broader population's adoption and perception of LLMs for everyday learning at a large scale, particularly in informal contexts where users voluntarily turn to LLMs for context-agnostic knowledge acquisition. This \textit{at-scale} understanding is important to rank current issues and benefits in order to provide a research agenda and actionable insights with the aim of advancing the design and applicability of LLMs as informal learning assistants towards a universally accepted, usable, and efficient learning tool. 

We address the identified gap by means of the following four research questions:
\vspace{-5mm}
\summarize{
\paragraph{\textbf{RQ1}} What demographic factors distinguish individuals who adopt LLMs for everyday learning from those who do not?
\paragraph{\textbf{RQ2}} What key factors motivate or discourage individuals from using LLMs for everyday learning? 
\paragraph{\textbf{RQ3}} What classes of LLM learners emerge regarding how, when, and in what contexts they integrate LLMs into their everyday learning practices?
\paragraph{\textbf{RQ4}} What are the attitudes of the different LLM classes regarding the learning effectiveness of these tools, their privacy implications, and the potential for overreliance in everyday learning? 
}

\vspace{5mm}
As such, our goal is to complement the listed related works by providing a broad, comprehensive, mixed-method inspection into LLM learners' demographics, interaction patterns, and perceptions of LLM-based informal learning  \textit{at scale}. To achieve this goal, we conducted a large-scale user survey over Prolific in February 2025 with 776 participants, including non-LLM adopters, LLM-adopters (but no learners), and LLM learners. We report on descriptive and inferential statistics between participants' demographics, the learning patterns participants employ with LLMs in everyday informal scenarios, and their diverse attitudes towards LLM learning at scale. To identify distinct learner profiles among LLM learners, we conducted a latent class analysis based on self-reported LLM usage for learning. Our analysis focused on uncovering unique patterns in the learning contexts where LLMs are employed, the specific tasks users perform, and the devices they utilize for learning. Finally, we examined how learners' attitudes regarding the tool's effectiveness, potential overreliance, and privacy implications in everyday learning scenarios relate to the identified learner types. 

88\% of participants adopted LLMs for learning. Learning adopters are younger, more educated, and predominantly male. We found the main driver for adoption and intensity of LLM use to be curiosity and openness to new technologies. On the other hand, the main barriers to adoption are mistrust in the accuracy and data handling of LLMs, as well as satisfaction with traditional learning methods. However, for those who use and learn with LLMs, the ability to receive personalized, on-demand support outweighed these concerns, with learners leveraging LLMs to enhance productivity and access just-in-time learning opportunities. We identified four distinct everyday learner profiles, namely \textit{Structured Knowledge Builders}, \textit{Self-Guided Explorers}, \textit{Analytical Problem Solvers}, and \textit{Adaptive Power Users}. Each learning type represents distinct patterns of engagement with LLMs -- from academic knowledge building to informal, mobile-first learning and broad, cross-contextual use.
Ultimately, LLMs are becoming a central tool in everyday life and are increasingly perceived as the new normal for improving learning and productivity.

\summarize{
\paragraph{\textbf{Contribution Statement}}

We contribute with 
1) an empirical account of how diverse demographic groups and learning styles incorporate LLMs as informal learning aids, 
2) a mixed-methods analysis of the benefits and challenges of using LLMs for everyday learning, and 
3) a set of actionable guidelines for designing and implementing AI-powered tools to support informal learning.

}

\section{Related Work}

In this section, we review informal learning, the perception and integration of LLMs in educational settings, HCI perspectives on conversational learning tools, and the challenges posed by these emerging technologies. 

\subsection{Informal and Everyday Learning}

There are many definitions and understandings of informal learning \cite{Eshach2007bridging}, yet in its simplest form, it is learning that rather happens in settings besides formal educational institutions (e.g., schools) \cite{Rogoff2016informal, marsick1999nature, manuti2015formal, livingstone2006informal}. As two opposing ends of a continuous learning spectrum \cite{Folkestad2006informal}, informal learning differs from formal education in several key ways. Unlike formal education, informal learning is non-didactic, driven by the learner's intrinsic motivation \cite{Rogoff2016informal}, and occurs in environments where learning is not explicitly organized or assessed \cite{malcolm2003interrelationships} but instead happens organically as part of daily life \cite{manuti2015formal}, such as family interactions, work settings, and community engagements \cite{Rogoff2016informal, Callanan2011Informal, evans2020informal}. The process of informal learning may involve self-directed, intentional learning, incidental learning, or socialization \cite{schugurensky2000forms}, being highly sensitive to contextual factors \cite{Cerasoli2018Antecedents, Folkestad2006informal}. Some scholars argue that everyday learning, i.e., learning with a heightened awareness of the surrounding context, coupled with a \quotes{culture of collaboration and trust}, can improve learning outcomes \cite{marsick1999nature} -- in particular for adults, as unstructured learning forms, such as informal and everyday learning, become their prevalent form of learning \cite{livingstone2006informal}. 
As AI and LLMs expand access to information in a more natural and everyday-fitting way, they can help people transform virtually every moment into a learning opportunity with context-appropriate timing and content. Unlike traditional search engines, which require learners to sift through and interpret multiple sources, LLMs can act as personal tutors that, e.g., explain complex concepts, adapt to the learner's prior knowledge, or answer specific questions in a conversational way. In this context, our work seeks to understand how modern LLMs contribute to this everyday learning process and where they might fill gaps left by current (or traditional) analogue and digital methods.

\subsection{Demographical Differences in LLM Use}
Large language models are used differently by individuals with different demographic characteristics, such as age, gender, educational background and personality traits\cite{jakesch2022groups}. Younger individuals are more keen to adopt and use LLMs \cite{Stein2024attitudes, kacperski2024characteristicschatgptusersgermany}, with those of high levels of agreeableness tending to have more positive views of LLMs, while older individuals often exhibit more skepticism \cite{Stein2024attitudes}. Furthermore, a clear gender gap exists in LLM adoption, with male users significantly outnumbering female users, both 4 months after the release of ChatGPT in November 2022 \cite{draxler2023gender}, as well as 30 months after\cite{sublime2024chatgpt}. This gap is particularly pronounced among younger age groups and diminishes as individuals age \cite{draxler2023gender}, with male users also reporting a higher frequency of LLM use compared to their female counterparts \cite{sublime2024chatgpt, draxler2023gender}. Similar numbers run for revisiting ChatGPT-generated content \cite{sublime2024chatgpt}. Among the group of researchers, women and non-binary researchers, along with those with more years of research experience, often express greater ethical concerns about LLMs \cite{morris2023scientists}. However, \citet{draxler2023gender} report the gender effect in general to be less pronounced among those who possess technical degrees.

LLM usage patterns also vary by expertise, with professionals and experts more likely to use LLMs for complex tasks such as writing formal texts or coding, while novices often explore the tool for entertainment or casual learning. Non-users often cite a lack of knowledge, ethical concerns, and the impersonal nature of LLM interactions as reasons for not adopting these tools \cite{draxler2023gender}.

In addition to gender, age, and expertise, other demographic factors such as race and nationality also influence LLM usage. Junior researchers and those who are non-White or non-native English speakers tend to use LLMs more frequently and report higher perceived benefits and lower risks \cite{liao2024llmsresearchtoolslarge}. As a consequence, LLMs can help address structural barriers faced by these groups in research, potentially promoting greater equity in academic environments \cite{liao2024llmsresearchtoolslarge}. Furthermore, researchers in computer science show greater comfort with disclosing their use of LLMs and express fewer ethical concerns compared to those in other disciplines \cite{liao2024llmsresearchtoolslarge}, suggesting disciplinary disparities in LLM adoption. 

However, there remains a gap in understanding how LLMs influence informal and everyday learning experiences across different demographic groups. 
We thus conducted a large-scale survey to capture insights from general-public learners, aiming to uncover the distinct and potentially varying effects of LLMs on learners' everyday learning habits. 

\subsection{LLMs for Learning}

\subsubsection{Early Perceptions}
A recent HCI review \citet{pang2025llmification} found that 15\% of the manuscripts published at CHI, which are focused on LLMs, are centred on \textit{education}. The interest in the influence of LLMs on education was indeed sparked early on. 
After the immediate release of ChatGPT, many scholars provided discussions reflected on the opportunities and challenges of LLMs both in general, as well as specifically in education and learning. \citet{DWIVEDI2023opinion} and \citet{bahrini2023applications} synthesize the existing narrative to highlight both the potential and threats of generative AI (GenAI), such as ChatGPT. As such, LLMs can enhance productivity across multiple sectors, such as business, healthcare, or science \cite{bahrini2023applications}, with in particular education and learning to \quotes{experience some of the most transformative impacts} \cite{DWIVEDI2023opinion}. \citet{DWIVEDI2023opinion} further emphasize biases, outdated training data, and misinformation as threats. \citet{kasneci2023chatgpt} in particular focused on the field of education, addressing both student and teacher perspectives on the opportunities and challenges of LLMs. In their opinion, students across all phases of education and learning (including informal and professional learning) can benefit from (personalized) assistance in writing, understanding complex concepts, encouraging critical-thinking and reflection skills, or developing certain language skills, among others. Yet, the authors argue that LLMs pose a series of challenges: copyright, overreliance, response shallowness, bias, privacy or data fabrication, and sustainability are among the high-profile ones. \citet{lin2024exploring} discussed ChatGPT's role in self-guided learning environments, listing its potential as a digital learning process companion for adult learners, though several concerns about over-dependence, unclear AI usage policies, and data updates to reduce incorrect outputs were raised. As LLMs evolve and transform into large reasoning models (LRMs), we can anticipate the emergence of more powerful capabilities such as self-reflection, along with more potential harms such as manipulation and hacking \cite{berti2025emergent}.

\subsubsection{Early Adopters}
Following the initial boom of LLMs, several studies have emerged examining their ongoing use with early adopters and their potential implications in education and learning. Three months after the ChatGPT release, \citet{hosseini2023exploratory} found that 40\% of their 420 participants had interacted with ChatGPT. Participants found research to profit more from LLMs than the fields of education or healthcare, citing a lack of depth and critical reasoning in medical education as the main challenge of ChatGPT. \citet{skjuve2023uxchatgpt} ran a qualitative questionnaire study among 194 early adopters about good and bad user experiences with ChatGPT. Their results reveal high pragmatic (e.g., helpful and comprehensive information, task simplification) and hedonic attributes (e.g., the \quotes{wow-factor}, entertainment, creativity) even in the early distribution stages of ChatGPT. Yet, the authors highlight the issue of LLMs being \quotes{convincingly wrong} as a critical challenge that HCI must address in order to ensure more reliable and positive future use. 
\citet{Albadarin2024review} conduct a systematic literature review of 14 early empirical studies (50\% qualitative, 20\% quantitative, and 29\% mixed-methods, with only two studies exceeding the participants' count of 100) to synthesize ChatGPT use patterns in educational settings. Negative consequences include risks such as misinformation, the digital divide, academic dishonesty, and, in extreme cases, manipulation. In the context of learning, these issues primarily manifest as a decline in critical thinking and problem-solving skills, along with reduced engagement in deep learning. The review also highlights that the use of ChatGPT worsens the lack of meaningful social interaction in group learning settings. On a more positive note, the authors highlight several beneficial ways in which students, particularly in higher education, utilize ChatGPT: as a writing and language support tool, a virtual on-demand assistant, an enhancer of self-directed learning, and a facilitator of complex concepts. 

In contrast to categorizing ChatGPT by learning tasks, \citet{stojanov2024latent} analyzed the types of learners, or \textit{reliers}, that emerge from relying on ChatGPT across 13 learning tasks. The study's analysis of 490 students resulted in five relier types: almost 40\% were \textit{versatile low reliers} with minimal dependence on the tool's various tasks; 16.5\% were \textit{knowledge seekers}, employing ChatGPT for content acquisition and summarization; 11.8\% were \textit{proactive learners} relying for feedback and planning; and 23.1\% were \textit{assignment delegators} who used it for drafting assignments and homework.
\citet{Tlili2023devil} qualitatively examined the use of ChatGPT in three waves -- a social media posts' analysis, interviews, and a UX study -- revealing initial optimism about ChatGPT in the space of social media with a grain of caution reported by some users. The further two study waves confirmed the dichotomy of perceptions around ChatGPT, framing ChatGPT as \quotes{devil and angel} in education. Similarly, \citet{Futterer2023global} discuss findings from a  social media analysis on both the global, general acceptance, as well as the acceptance in education, of early ChatGPT users -- unfolding mixed sentiments regarding education. A further social media analysis \cite{HADIMOGAVI2024chatgptBlessing} related specifically to educational contexts revealed that ChatGPT's role spans various educational contexts, including content creation in higher education, language learning in K-12, or problem-solving assistance in practical skills training. Despite the dominantly positive view among adult professionals, the study found that negative concerns, such as academic dishonesty, misinformation, and privacy issues, to be in particular relevance for K-12 and higher education contexts.  
In terms of specifically student use in higher education, \citet{NILOY2024why} identified several factors influencing the use of ChatGPT in higher education. The study found that factors such as ease of access, time-saving benefits, content inseparability, aided learning, and cognitive miserliness significantly influenced students' intention to use ChatGPT -- at the same time contradicting previous research works (e.g., \cite{DWIVEDI2023opinion}) by showing that technical knowledge of ChatGPT had no significant impact. \citet{chan2023expectancy} also examined the relationship between student perceptions and their intention to use GenAI, finding that perceived value had a strong positive correlation with usage intentions, while perceived cost had a weak negative correlation. In contrast, a recent user survey \cite{almurshidi2024understanding} on 366 students' perceptions of GenAI suggested that knowledge on both the benefits \textit{and} limitations of LLMs positively correlates with students' willingness to use GenAI in education. In a comparative study of students from pure and applied science fields, \citet{qu2024disciplinary} found that students in applied academic disciplines (i.e., engineering and science students) had greater knowledge and adoption willingness of GenAI, particularly for cognitively demanding tasks.

\subsubsection{Interaction Patterns of LLM Adopters}
Regarding concrete uses of ChatGPT in self-directed learning for writing, a recent user study with 384 students found that postsecondary students primarily used ChatGPT for brainstorming and idea generation in writing tasks \cite{wang2024understanding}. The study identified a use trajectory where students, initially motivated by curiosity or the need to complete tasks, continue using the tool due to the perceived task-benefits gained from their initial experiences, with the tasks including writing capabilities, information variety, or perceived usefulness, among others \cite{tiwari2024drives}. A further large-scale user survey \cite{strzelecki2024use} with 534 students revealed that established routines, perceived performance enhancement, and hedonic motivation are among the main drivers of ChatGPT use intention. 

In the study by \citet{wang2024understanding}, participants reported a strong sense of responsibility for their learning by cross-checking the output generated by ChatGPT. Ultimately, participants were divided on whether using ChatGPT improved their writing skills: some felt encouraged to practice writing and revising more frequently, while others felt the need for longer use to make an informed judgment. Similarly, for the task of writing, \citet{ozcelik2024cultivating} found that, in a much smaller sample of 11 participants, ChatGPT is an effective tool for helping students to develop their formal writing skills, particularly with regard to self-editing. However, they highlight the need for further functional improvements to enhance ChatGPT's effectiveness. 

To understand concrete use patterns of how and how much students use LLMs, \citet{sublime2024chatgpt} surveyed 395 students from as early as 13 years of age to university level, further investigating the influence of demographic differences on usage patterns. The same study investigates \textit{non-adopters} too, aiming to illuminate their reasons for non-use. The study revealed a number of interesting findings. First and foremost, LLMs have reached an audience of at least 70\% across all examined age groups and education levels, with an increasing use the further students are in their education level. Students reach hand to LLMs mostly on their computers \textit{and} smartphones, and report high satisfaction with the output of LLMs. 2/3 of students aged between 13 and 16 use LLMs, however, only 20\% revise the output. This fraction increases with the age of students (up to 50\%). The authors thus highlight the necessity of teaching critical evaluation and reflection on the output the AI generates -- a skill older students in their participants' possessed, probably due to the lack of availability of AI in their formative educational years. Non-users made up 13\% of the sample of participants, with perceived inappropriateness and academic honesty cited as the most common reasons for avoidance. Students reported using ChatGPT additionally for document writing (nearly 75\% of its usage). 

Informal and implicit learning in the workplace has been recognized as a key factor in professional development and the work environment \cite{Eraut2004informalworkplace}. Focusing on researchers and academics, \citet{liao2024llmsresearchtoolslarge} conducted a survey with 816 participants to explore current usage patterns of LLMs in their work pipeline, as well as their perceptions of the risks, benefits, acceptability, and demographic differences related to LLM use in research. 
Their findings indicate that 81\% of researchers have already integrated LLMs into some part of their workflow, primarily for tasks related to paper preparation. The most common uses include information seeking (e.g., discovering related papers or topics, aiding in summarization and explanations, or answering factual questions) and editing. Additionally, computer science researchers are more open about disclosing their use of LLMs, expressing fewer ethical concerns. This trend is also observed among non-white, non-native English speakers. In general, the issue (or opportunity) of equity emerged as a significant topic of discussion among the study's participants. Moreover, participants were more concerned about the LLM use of others than their own, indicating a \quotes{third-person effect} \cite{davison1983third}.

\summarize{\paragraph{\textbf{Research Gap}}Unlike most of the studies listed, our work focuses on the active, real-world use of LLMs in informal, everyday learning at scale. We aim to gather insights from a broad and diverse range of users, without imposing any specific recruitment criteria (e.g., profession, age, or gender), drawing from their lived experiences since the widespread release of LLMs in November 2022. Our research is guided by key questions about who is using LLMs, when and where they are being used, how they are being applied, and what is being learned in these everyday contexts. Additionally, we cross-examine these experiences with participants' perceptions of how LLMs have impacted their learning habits, particularly in terms of learning efficiency, reliance on LLMs, and privacy concerns.}

\section{Methodology}

\subsection{Survey Design}

Our survey was developed using an approach that integrated an examination of current scientific literature, triangulated with a GPT-generated questionnaire draft.
We began by reviewing existing literature on the role and use of LLMs and similar AI tools in education and learning. This initial step allowed us to identify several key constructs, such as the general and specifically learning tasks performed with LLMs (e.g., \cite{bodonhelyi2024userintentrecognitionsatisfaction, stojanov2024latent}), the diverse learning contexts in which these models are used (e.g., \cite{HADIMOGAVI2024chatgptBlessing}) or the benefits and drawbacks of employing LLMs(e.g., \cite{skjuve2023uxchatgpt}  and specifically for learning \cite{HADIMOGAVI2024chatgptBlessing}). Our objective was to complement the listed works' findings by quantifying these qualitatively established constructs through our survey with the general public. 
Inspired by the more recent work of \citet{sublime2024chatgpt}, we employed branching logic to include questions about general LLM non-use and non-use specifically in the domain of learning. In these branches, we were interested in reasons for non-use and the demographic attributes of non LLM users.  
Our review of related literature also revealed additional constructs emerging from \textit{actual LLM usage}, that is, beyond initial (hypothetical) perceptions around LLMs. Consequently, we added items addressing the interaction modalities employed with LLMs, the device and time context of LLM use, payment for LLM services (informed by frameworks such as those in \cite{HAMARI2020whypay, Hamari2017Service}), changes in learning habits and perceived learning effectiveness, as well as implemented privacy measures and (over-)reliance effects \cite{hamdan2023impact} of LLMs on learning.
With the built knowledge of existing and missing constructs in the literature, we prompted ChatGPT to create an initial draft of a questionnaire to examine \quotes{who uses how LLMs for educative and learning purposes} in January 2025. 
We subsequently refined the prompt to add questions on the identified missing constructs.  
We curated the resulting question set and answer options in several waves, based on existing literature and pilot testing, with emphasis on the categorical multiple-choice answer options. 
As a result of the described approach, the survey first assessed participants' prior engagement with AI tools and multimodal LLMs by asking them to indicate which LLM tools they had used. Those who had never used such tools were routed to blocks exploring reasons for non-use, while LLM users answered follow-up items on interaction modalities (e.g., text-to-text, image-to-text) and LLM payment options. LLM users were then asked whether they use LLMs for learning, too -- if not, the survey listed multiple options to choose from as reasons not to use LLMs for learning. Otherwise, the survey branched to specific tasks performed with LLMs \textit{for learning}.  
The survey then probed the learning contexts in which LLMs are applied (from \cite{HADIMOGAVI2024chatgptBlessing}), capturing the frequency and duration of use (adapted from \cite{ess2025dataportal}), triggers for adoption, devices and platforms, social settings of use as well as the range of inquiries \cite{bodonhelyi2024userintentrecognitionsatisfaction}  and learning tasks (such as seeking factual information, brainstorming, and problem-solving, compiled from \cite{sublime2024chatgpt, liao2024llmsresearchtoolslarge, stojanov2024latent}) through both categorical and continuous measures. Respondents' perceptions of the benefits and drawbacks of using LLMs for learning were evaluated with a set of multiple-choice questions, followed by Likert-scale items focusing on the LLM's critical thinking impacts and over-reliance, and challenges such as misinformation or usability issues. The survey also explored changes in learning habits and data privacy concerns to contextualize the impact of LLMs on everyday learning practices. 
To conclude the section on the use of LLMs for learning, the questionnaire explored potential improvements for LLMs in educational contexts and assessed participants' willingness to recommend them to others. 

Finally, for a more comprehensive demographic analysis, we included five single-choice questions on AI literacy  \cite{HORNBERGER2023AILiteracy} that assessed participants’ understanding of AI functionalities and limitations, the Affinity for Technology Interaction (ATI) scale\cite{Franke2019ATIscale}, and a short Big-5 personality questionnaire \cite{RAMMSTEDT2007big5}. 

\subsection{Procedure \& Participants}

We implemented the questionnaire using Qualtrics and recruited participants from Germany using Prolific\footnote{\url{www.prolific.com} [date accessed: \today]} in several batches during the month of February 2025. The recruitment text stated that we are looking into how \quotes{multimodal large language models (LLMs) [are used] in your everyday life for learning and productivity. Even if you do not use LLMs, for learning or in general, we would still be interested to hear your reasons for not using them.} 

Participants had to possess fluent English skills. Besides an equal gender distribution, no other screening criteria were employed. This was to ensure a wide range of participants' attributes. The questionnaire began by listing the GDPR-conform consent form to which participants had to agree in order to continue with the questionnaire.

818 participants started the questionnaire beyond the consent point, with 801 participants completing it within the designated time of 55 minutes. We excluded 25 participants after data inspection, as these have failed our attention check: these participants selected certain tasks to be completed with LLMs, with at least one task's frequency later specified as 'never used'. Our final participants' pool thus consists of $n=776$ participants, aged 31.6 years on average ($SD=9.87$, $max=72$). The gender distribution of the participants was nearly balanced, with males representing 49.9\%, females 48.5\%, and non-binary participants making up 1\%. Participants needed on average 19.74 minutes ($SD=11.56$) to fill out the questionnaire. and received monetary compensation (12.82EUR/hour) via the platform after finishing. 

\subsection{Analysis}

\subsubsection{Latent Class Analysis}

To identify distinct learner profiles based on their interaction patterns with LLMs, we employed latent class analysis (LCA), a probabilistic modeling approach that identifies unobserved (latent) subgroups within a population based on patterns in categorical variables. LCA assumes that an underlying latent categorical variable accounts for the associations among observed variables, grouping individuals into discrete and mutually exclusive classes that share similar response patterns.

In this study, we applied LCA to three key dimensions of learning behavior: \textit{devices used} (e.g., desktop, laptop, smartphone, tablet), \textit{learning contexts} (e.g., higher education, K–12, lifelong learning, professional development), and \textit{learning tasks} (e.g., brainstorming, coding assistance, feedback seeking). Each of these categorical variables served as an indicator in the LCA model, allowing us to uncover latent learner profiles based on their behavioral patterns.

To determine the optimal number of latent classes, we fitted models with varying numbers of classes and evaluated model fit using two widely used information criteria: the Bayesian Information Criterion (BIC) and the Akaike Information Criterion (AIC). Both metrics penalize model complexity to prevent overfitting, with lower values indicating better model fit as depicted in Figure \ref{fig:lca-fit}. Based on these criteria, a four-class solution provided the best balance between explanatory power and parsimony, suggesting the existence of four distinct learner profiles. LCA was conducted using the poLCA package in R, which estimates model parameters using an expectation-maximization (EM) algorithm combined with Newton-Raphson optimization. The resulting class-specific conditional probabilities were examined to interpret the distinct behavioral patterns associated with each class.

\section{Results}

We present our findings organized around the four research questions. To address RQ1, we first examine demographic differences by dividing our participants' pool of 776 individuals into three groups based on LLM usage for learning, that we further address as:
\summarize{
    \paragraph{\textbf{Non-LLM Users}} \textbf{15} respondents (about 2\%) have never used any LLM or similar AI tool.
    \paragraph{\textbf{Non-LLM Learners}} \textbf{83} respondents (11\%) have used LLMs but not for everyday learning purposes.
    \paragraph{\textbf{LLM Learners}} \textbf{678} respondents (87\%) have incorporated LLMs into their everyday learning practices.

}

\vspace{5mm}
With this grouping, we make descriptive and inferential comparisons between those who adopt LLMs for learning versus those who avoid them (RQ1). We then perform a mixed-method analysis of triggers and hindrances for LLM adoption for learning to investigate RQ2. 
To address RQ3, we conduct an LCA among those who adopt LLMs for learning to uncover underlying patterns in timing, context, and manner of LLM usage for day-to-day learning. Concluding with RQ4, we run several regression tests to show how the resulting learner profiles stand in relation to perceived attitudes of effectiveness, reliance, and privacy with LLM use. We accompany these results with a qualitative analysis of the answer set on how LLMs changed participants' approaches and habits with learning. 

\subsection{RQ1: Demographic Differences}

Our analysis reveals several demographic trends between the three groups of participants. 

\begin{figure}[h]
    \centering
    \begin{subfigure}[b]{0.48\linewidth}
        \centering
        \includegraphics[width=\linewidth]{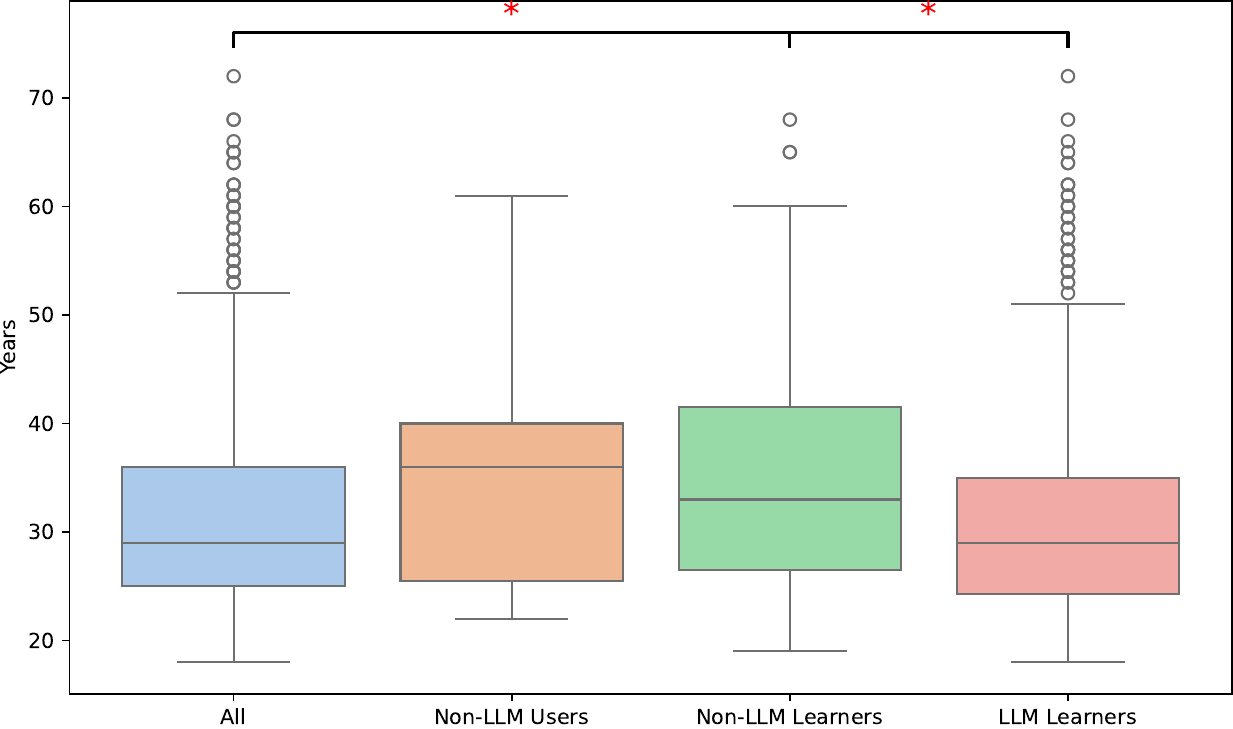}
        \caption{Age Distribution}
        \label{fig:age-distribution}
    \end{subfigure}\hfill
    \begin{subfigure}[b]{0.48\linewidth}
        \centering
        \includegraphics[width=\linewidth]{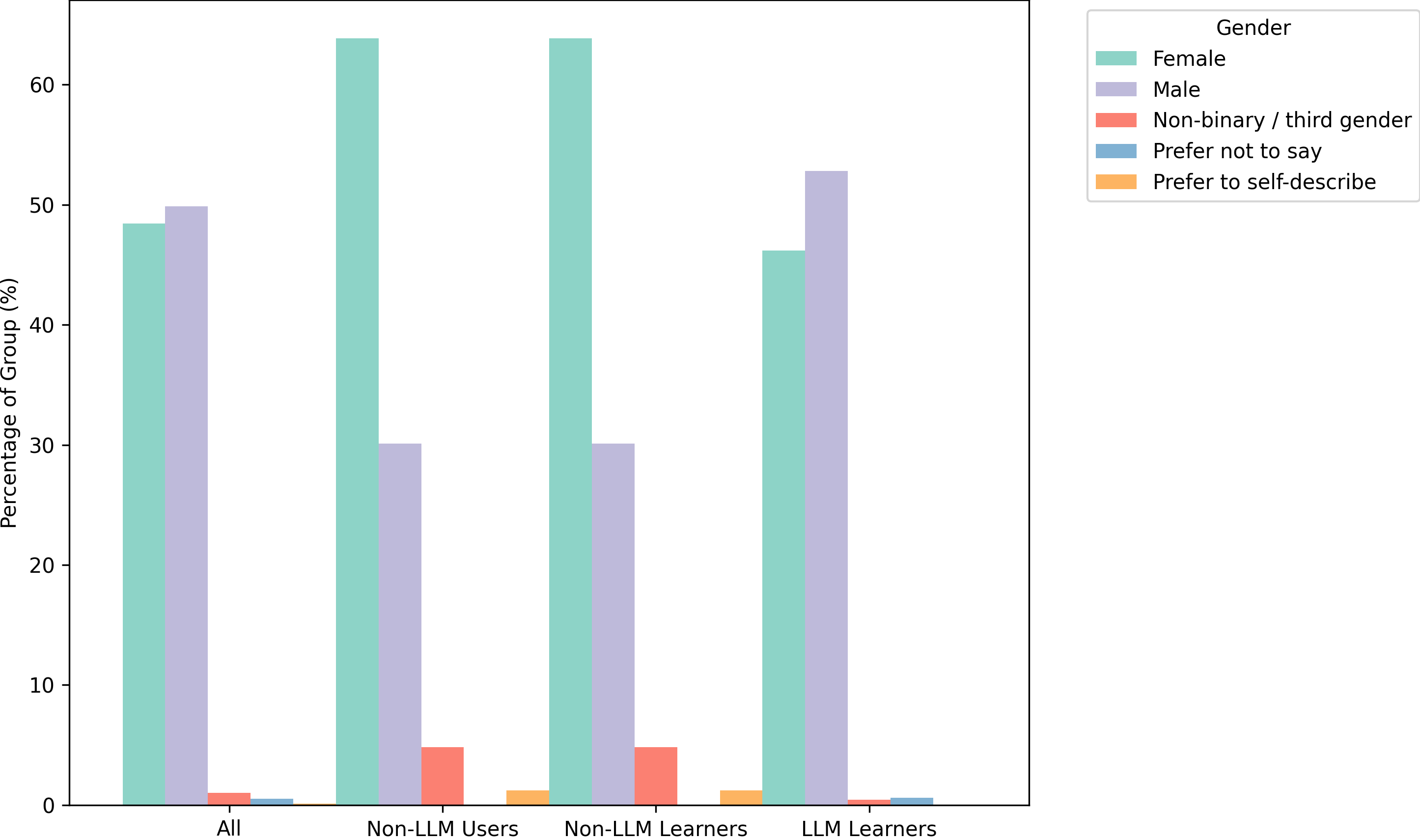}
        \caption{Gender Distribution}
        \label{fig:gender}
    \end{subfigure}
    \caption{Demographic distributions: (a) age and (b) gender}
    \label{fig:demographics}
\end{figure}

\paragraph{Age} LLM learners tend to be younger on average ($M=31$, $SD=9.57$), whereas those who did not use LLMs for learning were older ($M=35$, $SD=11$) on average, see \autoref{fig:age-distribution}. An ANOVA test confirmed a significant difference in age ($F=6.68$, $p<0.01$). In particular, the LLM learners were significantly younger than the group who use LLMs (but not for learning, $p= 0.004$). 

\paragraph{Gender} 
Among LLM learners, 53\% participants were male ($n=358$) and 46\% ($n=313$) female (see \autoref{fig:gender}), whereas in the group that avoided using LLMs for learning, women formed a majority with 64\% ($n=53$ female, $n=25$ male). A chi-square test found this difference significant ($\tilde{\chi}^2=15.5$, $p<0.001$), meaning that men in our participants' sample were more likely to embrace LLMs as learning tools than women.

\paragraph{Education Level}
Educational attainment was relatively high overall, but non-LLM users had lower education levels on average (albeit not siginificant). 40\% of non-users ($n=6$) hold at least a bachelor's degree, compared to 57–61\% ($n=386$ and $n=51$) of those who use LLMs (with or without learning, respectively). 

\begin{figure}[h]
    \centering
    \begin{subfigure}[b]{0.48\linewidth}
        \centering
     \includegraphics[width=\linewidth]{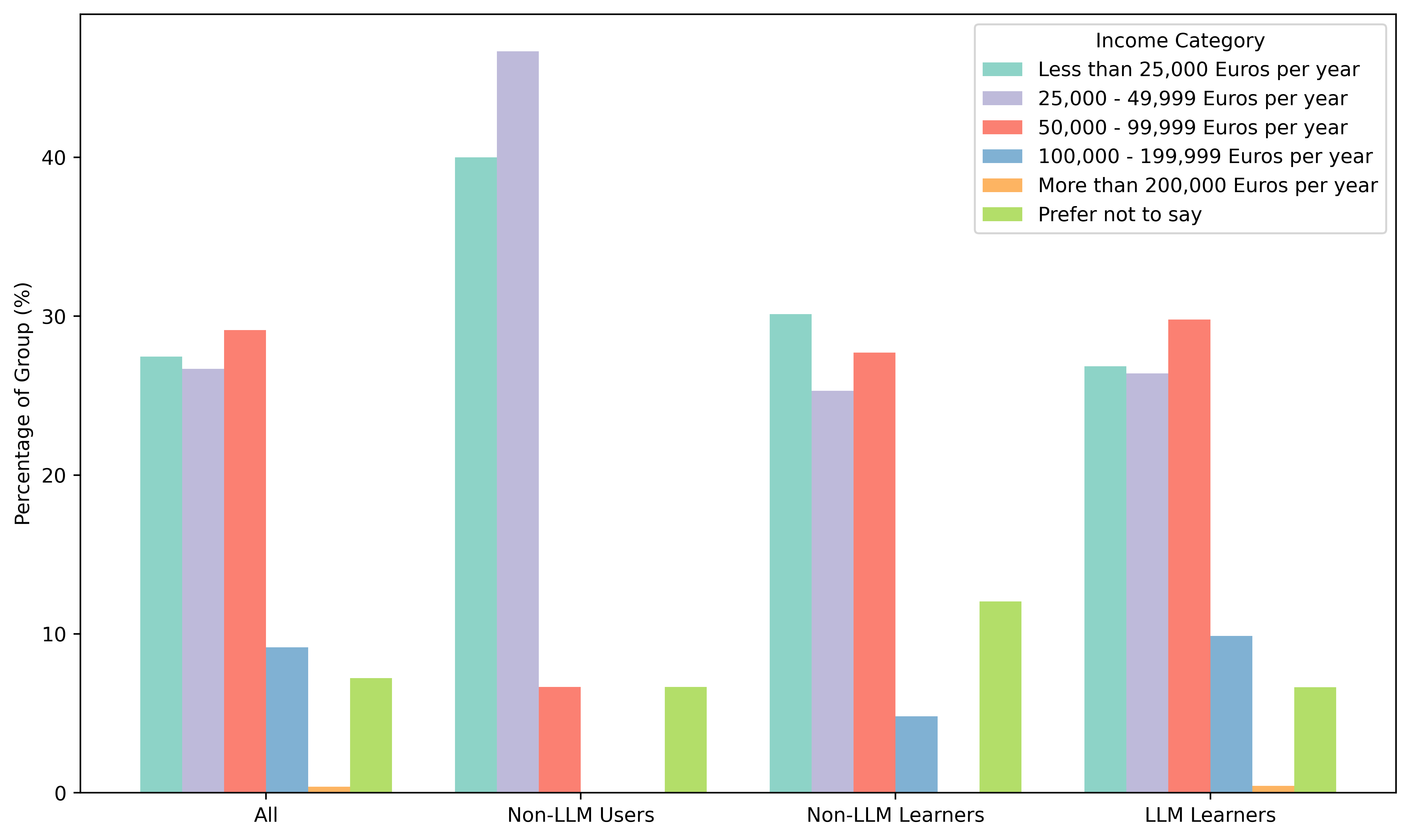}
    \caption{Income Comparison}
    \label{fig:income}
    \end{subfigure}\hfill
    \begin{subfigure}[b]{0.48\linewidth}
        \centering
        \includegraphics[width=\linewidth]{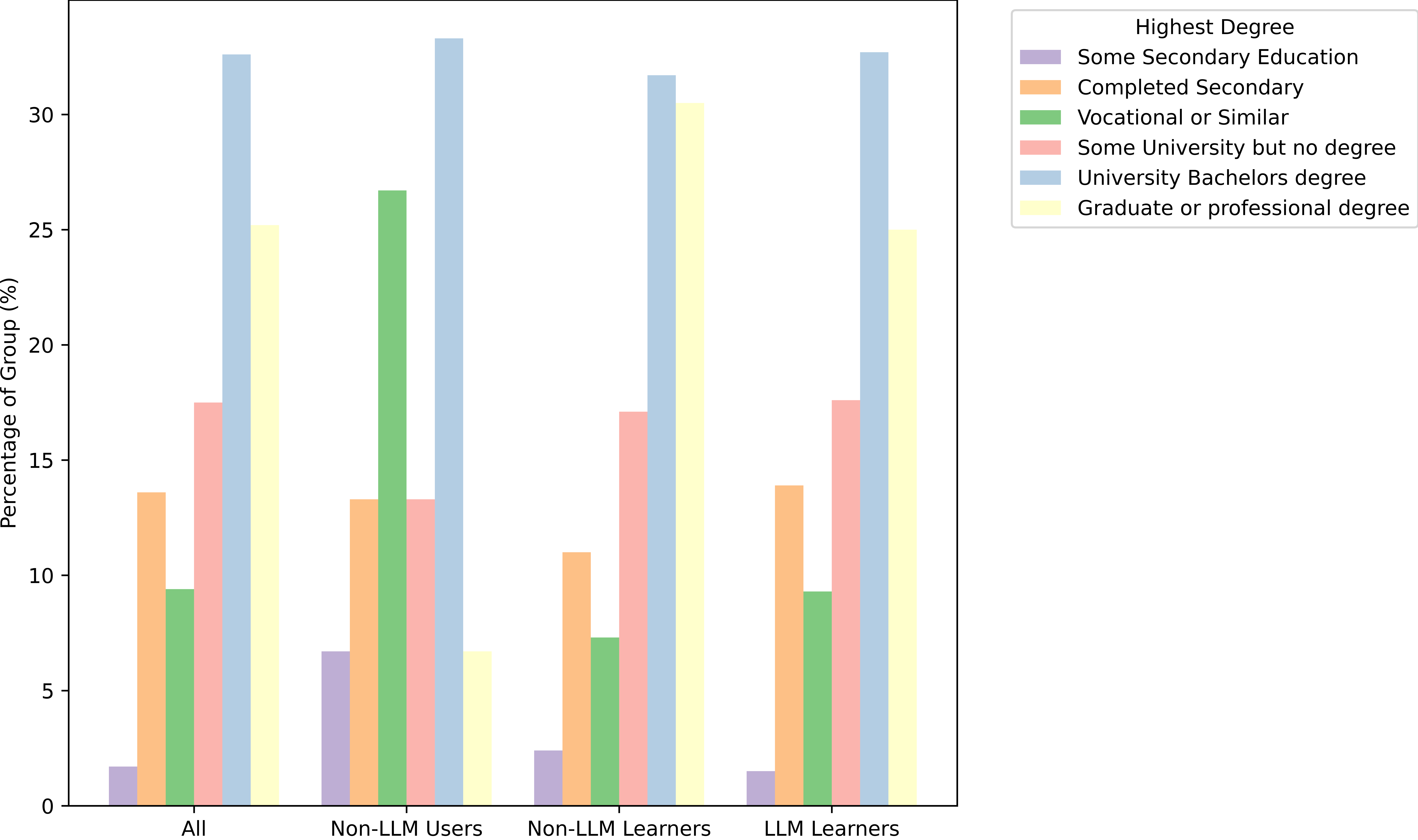}
        \caption{Educational Degree Comparison}
        \label{fig:education}
    \end{subfigure}
    \caption{Demographic distributions: (a) income and (b) education level}
    \label{fig:demographics-edu-inc}
\end{figure}

\paragraph{Income}
LLM learners were distributed across income levels, while non-users clustered in lower incomes. About 87\% ($n=13$) of non-users earned below EUR50K/year versus 53\% ($n=407$) of LLM users (learning or not) in that range. We observed more high-income individuals among learning-adopters – nearly 10\% ($n=67$) of them had incomes greater than EUR100K, whereas none of the non-users did. Nonetheless, adoption was not limited to any one income bracket; even within lower income groups, many participants were using LLMs for learning, as depicted in \autoref{fig:income}.

\paragraph{Technology Attitudes}
We measured an \textit{AI Literacy} score by means of five multiple-choice questions (single correct answer, score 0 to 5) from \cite{HORNBERGER2023AILiteracy}, next to the technology affinity score measured by \cite{Franke2019ATIscale}. 

\begin{figure}[t]
    \centering
    \begin{subfigure}[b]{0.48\linewidth}
        \centering
        \includegraphics[width=\linewidth]{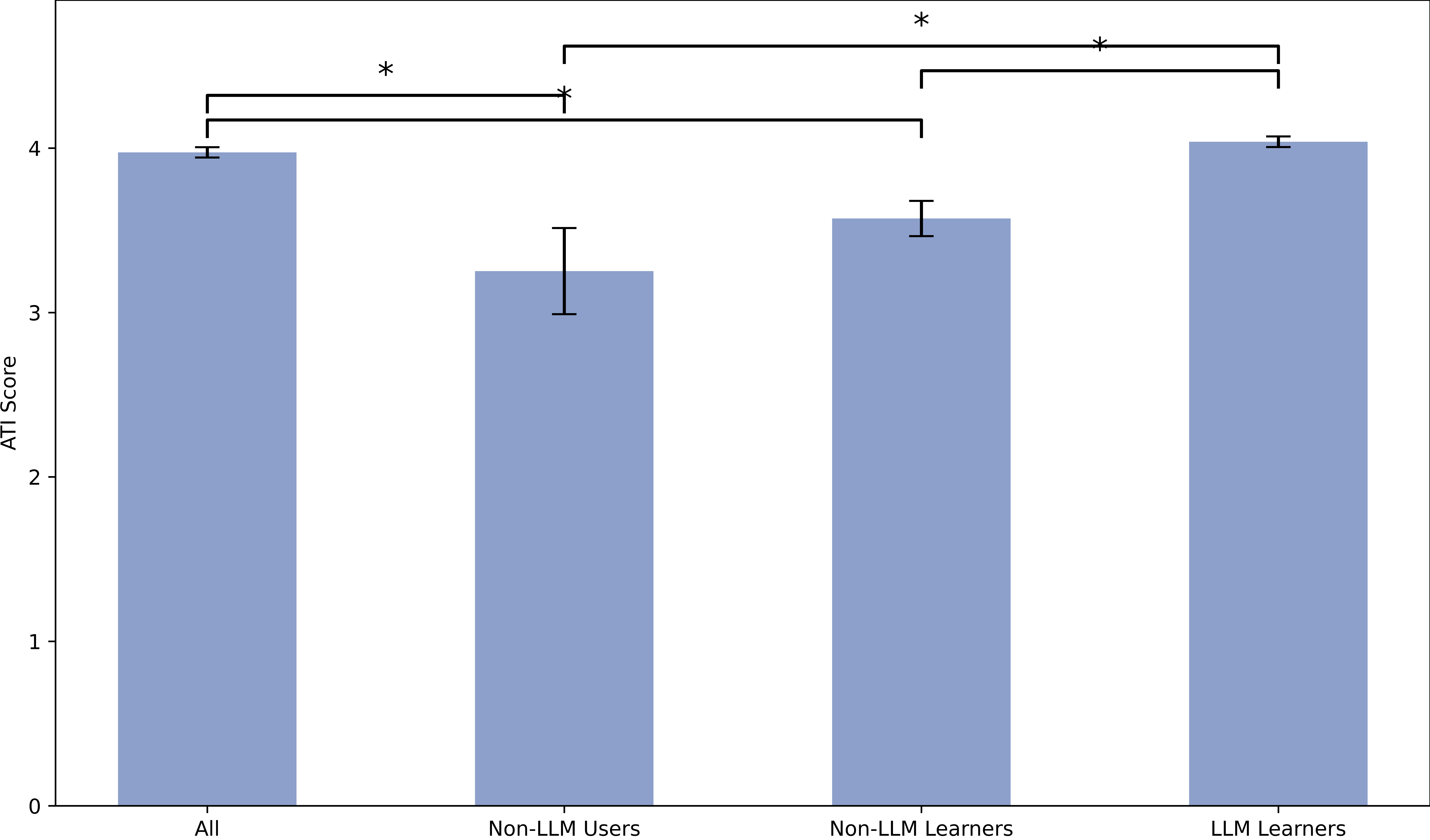}
        \caption{ATI Score Distribution}
        \label{fig:ati-score}
    \end{subfigure}\hfill
    \begin{subfigure}[b]{0.48\linewidth}
        \centering
        \includegraphics[width=\linewidth]{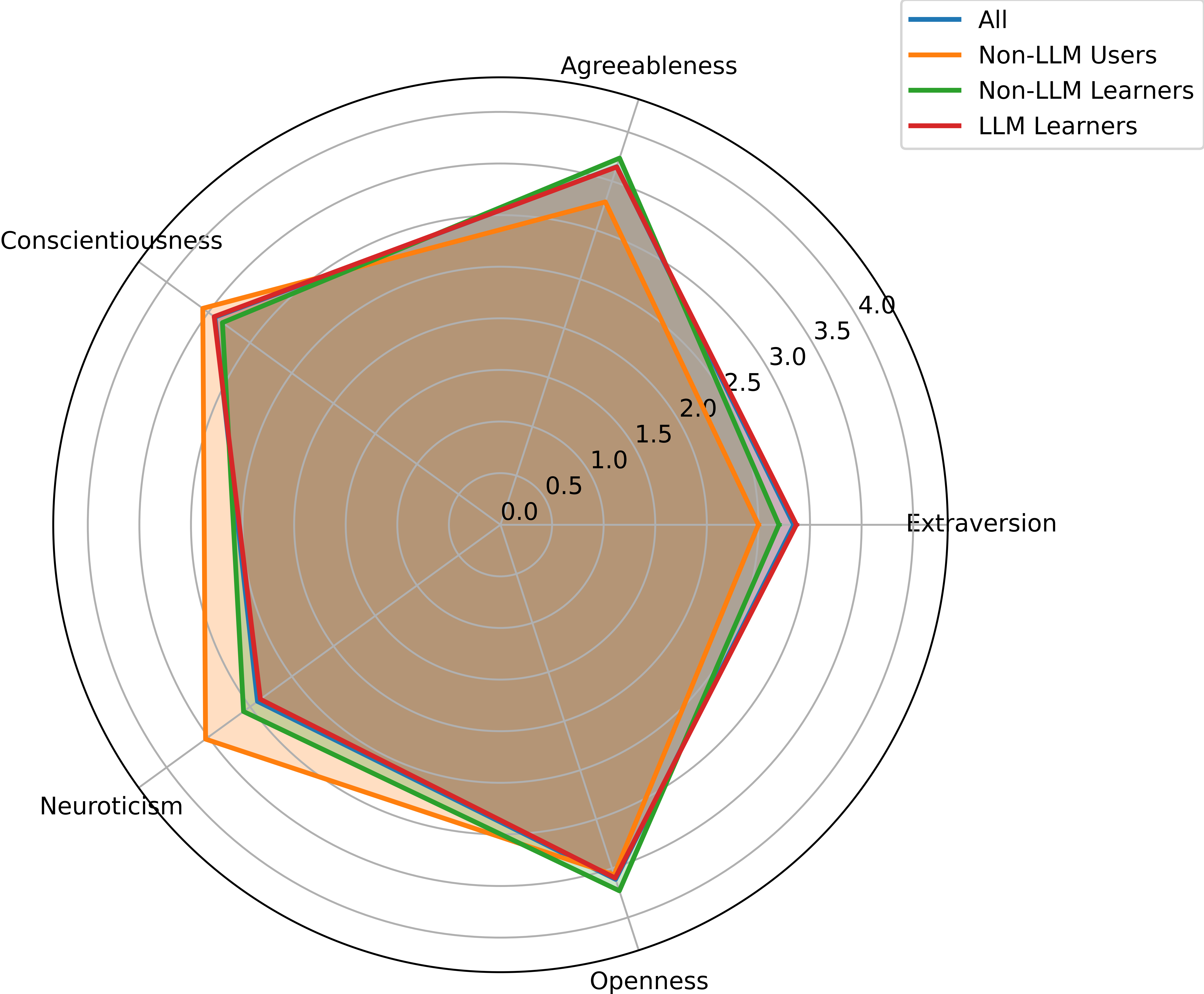}
        \caption{Personality Distribution}
        \label{fig:personality-spider}
    \end{subfigure}
    \caption{Distributions of ATI scores and personality traits: (a) ATI score distribution, (b) big-5 personality distribution}
    \label{fig:score-personality}
\end{figure}

Even those who have not adopted LLMs are reasonably aware of AI basics, as AI literacy scores did not differ strongly between groups ($\tilde{x}=3$ and $\tilde{x}=3.2$ for LLM learners and non-LLM learners, respectively, vs $\tilde{x}=2.87$ for non-LLM users). Attitude toward technology, however, did differ: LLM-learners scored highest on the ATI scale ($M=3.92$. $SD=0.55$, $\alpha=0.71$ out of 5), significantly above those who avoided using LLMs ($p<0.01$), as per \autoref{fig:ati-score}.

\paragraph{Big-5 Personality} Personality factors did not show large differences between groups, although non-users trended slightly higher on neuroticism and lower on agreeableness (likely due to the small non-user sample, these differences were not statistically significant), as depicted in \autoref{fig:personality-spider}.

\summarize{\paragraph{\textbf{RQ1 Main Findings}}

According to our findings, the adoption of LLMs for learning has reached a broad proportion of 88\% in our sample. Statistically, we found significant differences in age and gender distributions between LLM learners and non-LLM learners, as well as in tech attitude scores. Education and income showed observable trends (learners skewing higher) though the group differences were less stark after excluding the very small non-user group. These patterns reinforce that early adopters of LLMs for learning look similar to early adopters of other tech \cite{CAI2017gender} – \textbf{younger, educated, and male-skewed.}}

\subsection{RQ2: Discouraging and Motivating Factors to Use LLMs for Everyday Learning}

\paragraph{Discouraging Factors and Challenges}
\label{sec:challenges} The primary reason for LLM avoidance reported by the non-LLM users group pertains to mistrust ($n=6$, $40\%$) in the factual output of LLMs. 13\% ($n=2$) of participants mentioned concerns regarding the handling of data (privacy) during the training process. Ethical considerations also play a role (\quotes{It feels like cheating}), with some participants expressing concerns about the environmental impact of LLMs ($n=2$), while others express reservations about excessive reliance on AI tools ($n=2$). One participant articulated their concerns with the prospect of not wanting to become \quotes{lazy}. Additionally, two participants felt unable to utilize LLMs effectively. 

\begin{figure}[ht!]
    \centering
    \begin{subfigure}[b]{0.48\linewidth}
        \centering
        \includegraphics[width=\linewidth]{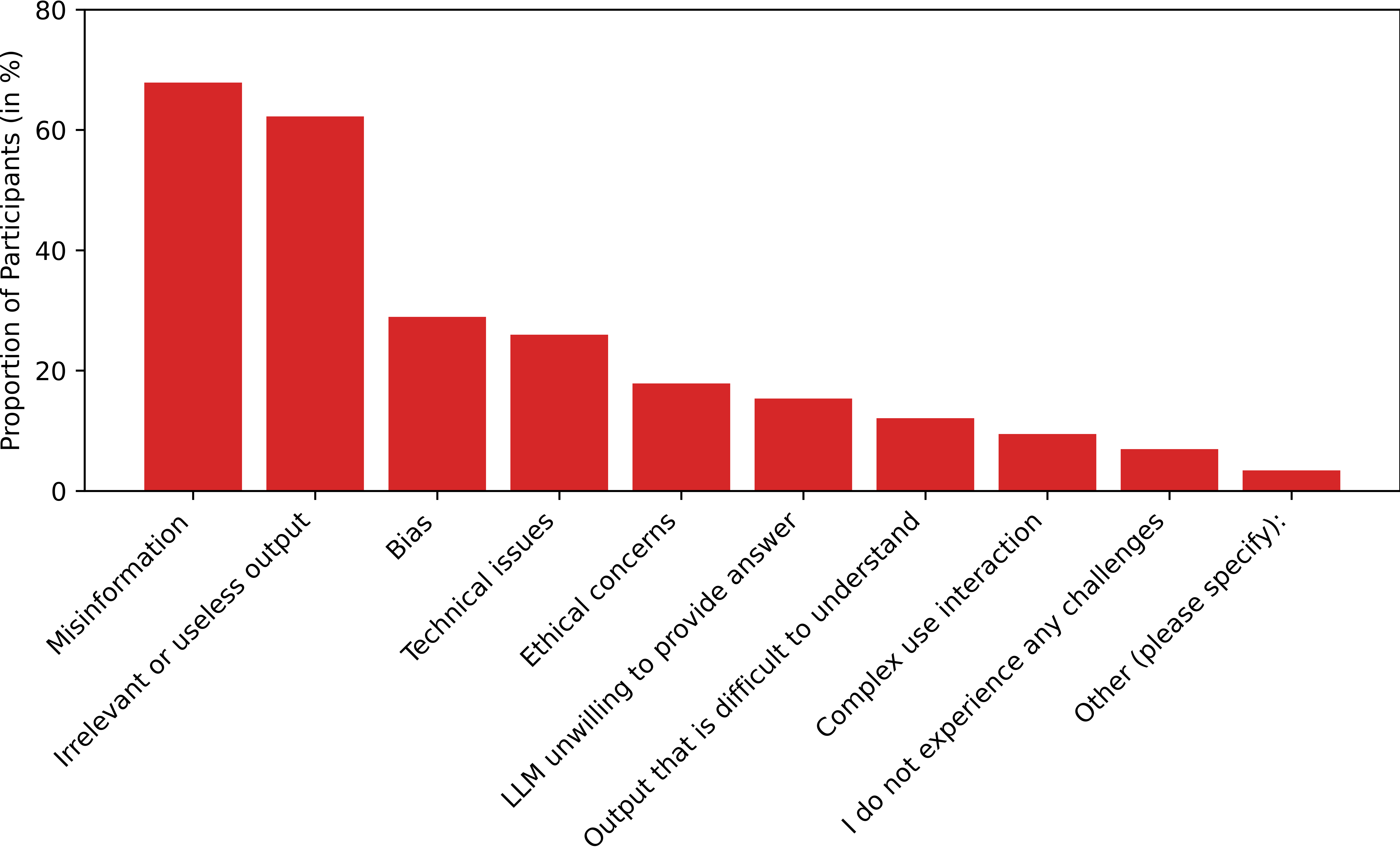}
        \caption{Challenges}
        \label{fig:challenges-learners}
    \end{subfigure}\hfill
    \begin{subfigure}[b]{0.48\linewidth}
        \centering
        \includegraphics[width=\linewidth]{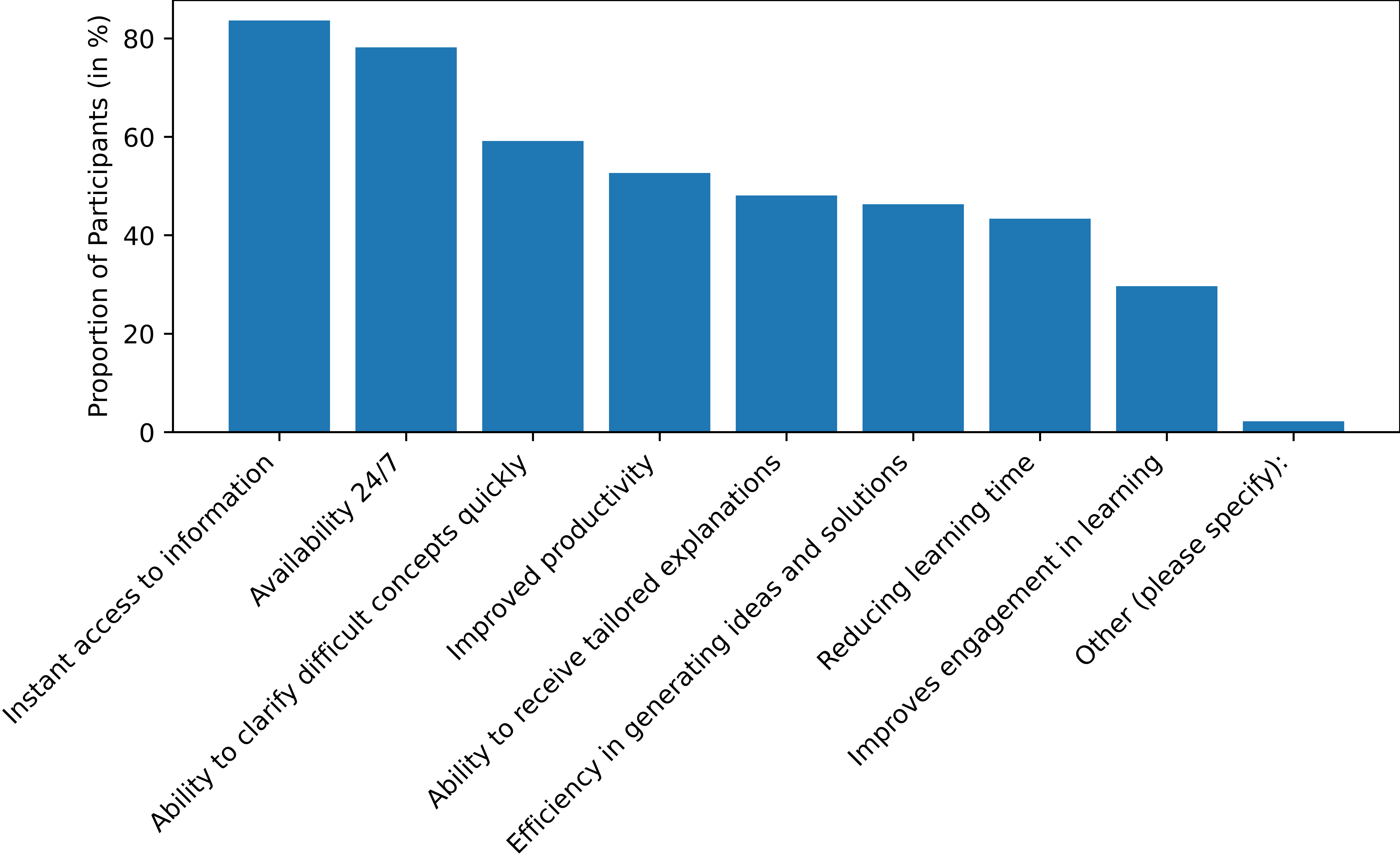}
        \caption{Benefits}
        \label{fig:benefits-learners}
    \end{subfigure}
    \caption{Proportion of reported (a) challenges and (b) benefits experienced by learners when using LLMs for learning}
    \label{fig:challenges-benefits}
\end{figure}

We further inquired with the group of non-LLM learners about their reasons for avoiding LLMs, which are more tailored to the use case of learning. Participants were permitted to select multiple reasons. These reasons exhibit overlap with those previously cited by non-LLM users. The predominant concerns ($n=52$) revolve around the reliability of factual output, ethical concerns ($n=25$) and data privacy ($n=19$). 

Conversely, 52\% non-LLM learners expressed a preference for traditional learning methods ($n=44$). Whereas almost a quarter ($n=20$) of participants cited a lack of motivation, one eight ($n=12$) of non-LLM learners indicated a perceived lack of skills necessary to effectively utilize LLMs for learning purposes. Notably, only two participants cited a strict prohibition by their educational institutions as a deterrent. The majority of non-LLM learners appear to adhere firmly to their stated positions of not using LLMs for learning, as 61\% ($n=51$) indicated that they have no intention of using LLMs for learning in the future, with 19\% ($n=16$) participants expressing some degree of hesitation. 

Among the potential factors that might encourage non-LLM learners to utilize LLMs for learning purposes, the responses were distributed between improved accuracy ($n=35$) and recommendations from AI experts and educators ($n=23$). Yet, a quarter ($n=21$) each selected \quotes{nothing} and \quotes{I stand firmly in not using LLMs for learning.}

LLM learners could specify challenges they face when learning with LLMs from a multiple-choice list. As \autoref{fig:challenges-learners} depicts, the most commonly selected challenge when interacting with LLMs was misinformation ($n=460$), accounting for 68\% of respondents. The second most frequent issue was irrelevant or useless output ($n=422$, participants or 62\%). Bias in AI-generated responses was noted in 29\% of cases ($n=196$), reflecting concerns about fairness and neutrality in AI outputs. Additionally, technical issues, such as system errors or response failures, affected 26\% ($n=176$). Ethical concerns were raised by 18\% ($n=121$), followed by 15\% ($n=104$) of LLM learners encountering instances where the language model was unwilling to provide an answer. The clarity of AI-generated output was selected in 12\% ($n=82$), with participants finding responses difficult to understand. Moreover, 9\% ($n=82$) mentioned experiencing complex user interactions, possibly indicating difficulties in effectively communicating with the model. Interestingly, 7\% ($n=47$) stated not experiencing any challenges and 3\% ($n=23$) reported facing other challenges that were not captured in the predefined categories.

\paragraph{Motivating Factors and Benefits}
We surveyed the LLM learners' group to list the motivating factors behind their adoption of these technologies. As per \autoref{fig:benefits-learners}, the major motivating factor listed was curiosity ($n=517$), followed by recommendations from social connections ($n=254$). In a similar vein, media coverage or news about LLMs was a trigger for 217 participants. In 157 cases, the educational institution or employer encouraged using LLMs for learning. 

85 participants shared how their use of LLMs for learning has changed over time and what makes them continue to use them. Many participants stated starting out of curiosity, which grew over time into exploring various capabilities of AI, which for some, led to increased prompting knowledge and knowledge on how LLMs operate in general: \quotes{I think they keep getting better and faster. I also have learned to use prompts better, so maybe that also helps.} Some participants were able to tunnel its use for \quotes{very specific tasks} -- others observe how the use of LLMs has spilled over into other various contexts of work and everyday chores: \quotes{First, I only learned for school, now I use it in my everyday life as well.}

We finally asked LLM learners to share their thoughts on the benefits of using LLMs for learning. Again, participants could choose from a multiple-choice list. The two top benefits revolve around access, responsiveness, and flexibility in unconstrained use: instant access to information emerged as the most widely recognized benefit ($n=567$, 83.63\% of LLM learners), closely followed by LLMs being available 24/7 ($n=530$, 78.17\%). 
59.4\% and 48\% of participants felt that LLMs were able to clarify difficult concepts quickly, at times with tailored explanations, respectively. Whereas a little more than half ($n=357$) of the participants stated to have experienced improved productivity, almost a third ($n=201$) experienced improved engagement in learning thanks to LLM use. Finally, 46.3\% ($n=314$) and 43.3\% ($n=294$) reported LLMs being efficient in generating new ideas and being helpful in reducing learning time, respectively.   
\summarize{\paragraph{\textbf{RQ2 Main Findings}}The primary barriers keeping people from using LLMs for learning are \textbf{lack of trust in the output accuracy and data handling} of AI, as well as satisfaction with traditional learning methods – essentially, \quotes{Why use this if I cannot fully trust it and I am doing fine with Google/books/teachers?}

Interestingly, the same concern of mistrust tops the chart of challenges among LLM learners, with their experiences showing that most have indeed encountered incorrect answers provided by the LLMs. Another insight is that some avoiders simply \textbf{do not know how to fit AI into their learning}, next to being concerned about the (non)ethical use of not becoming over-reliant or misusing AI in academic settings.

LLM learners emphasize the ability to get personalized explanations on any topic at any time. This lowers the barrier to entry for exploring new topics, given that curiosity was the biggest adoption driver. It's like having a tutor on call, albeit one that sometimes lies to you. For this group, the benefits clearly outweigh the challenges of LLMs.
}

\subsection{RQ3: Learning Tasks and Contexts}

For the 678 participants actively using LLMs in their learning, we explored which LLMs and their modalities they turn to for learning, next to what tasks they use LLMs for. Unsurprisingly, ChatGPT is by far the most commonly used LLM, with participants relying strongly on the text-to-text modality for a wide variety of tasks.   

Subsequently, we explore the when, where, and how these tools fit into participants' daily learning routines. The picture that emerges is one of flexible, self-directed usage: people integrate LLM assistance whenever they need it, largely on personal devices, and mostly on their own (i.e., not in groups). Below, we first break down the task usage patterns, followed by context patterns. We then describe the four distinct learner profiles identified by LCA.

\subsubsection{The What of Everyday Learning with LLMs}

\paragraph{Which LLMs?}
The median amount, per participant, of LLMs used for learning is one ($M=1.38$, $SD=0.72$, $max=6$). Among the chatbots used for learning, OpenAI ChatGPT was by far the most frequently mentioned, with 93\% ($n=632$) of LLM learners reporting its use. Google Gemini ($n=86$), Microsoft Copilot ($n=81$), and DeepSeek ($n=64$) follow next. A little less than 5\% ($n=33$) reported using LLMs outside of these platforms, such as Perplexity, Claude or Grok.

\paragraph{Inquiries and Learning Tasks}
\label{sec:tasks}

We asked participants to share which inquiries they pursue with LLMs, guided by the inquiry taxonomy by \cite{bodonhelyi2024userintentrecognitionsatisfaction}. This taxonomy encompasses 24 inquiries that evolve not only around learning tasks, but also self-reflective or artistic practices, among others. 

Participants reported conducting an average of 7.53 different inquiries ($SD = 4.40$) with LLMs, with the inquiry variance per person ranging from 1 to 24 ($\tilde{x} = 7$). This indicates a wide variability and understanding of the inquiries individuals pose with LLMs for daily learning-related activities.

\begin{figure}[h]
    \centering
    \includegraphics[width=\linewidth]{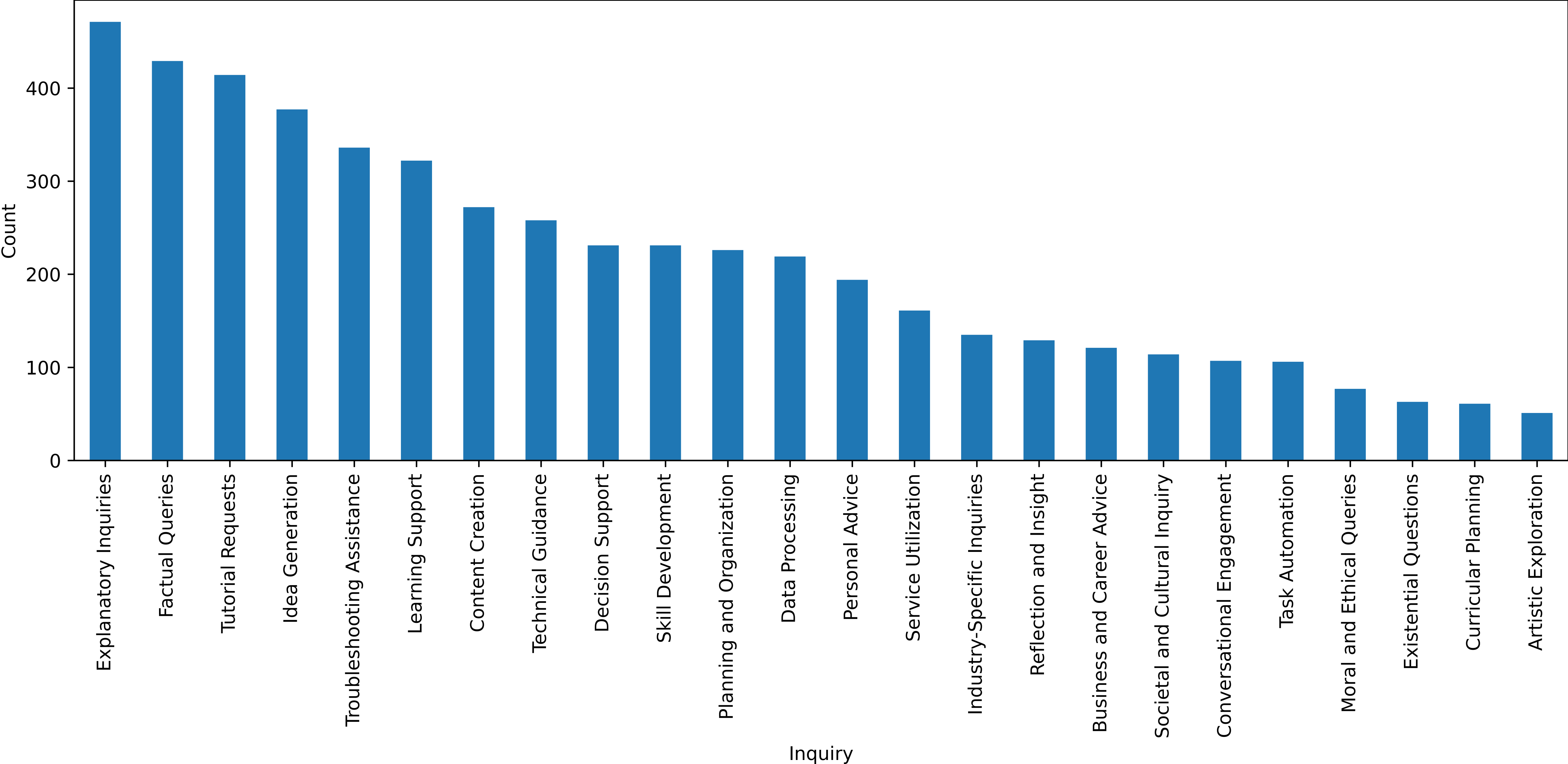}
    \caption{Frequency of mentioned inquiries}
    \label{fig:learning-inquiries}
\end{figure}

The most frequently performed tasks were \textit{Explanatory Inquiries} ($n=471$, 69.5\%), \textit{Factual Queries} ($n=429$, 63.3\%), and \textit{Tutorial Requests} ($n=414$, 61\%). These point to learners relying on LLMs for gaining explanations, retrieving factual information, and following step-by-step instructions. Next come \textit{Idea Generation} ($n=377$, 55.6\%) and \textit{Troubleshooting Assistance} ($n=336$, 49.6\%), indicating that LLMs play a role in creativity and problem-solving. Other frequently reported tasks included Learning Support ($n=322$, 47.5\%), \textit{Content Creation} ($n=272$, 40.1\%), and \textit{Technical Guidance} ($n=258$, 38\%). Less frequently mentioned (but still notable) were tasks related to \textit{Decision Support} ($n=231$, 34\%), \textit{Skill Development} ($n=231$, 34\%), \textit{Planning and Organization }($n=226$, 33.3\%), and \textit{Data Processing} ($n=219$, 32.3\%). Personal and professional guidance, such as \textit{Personal Advice} ($n=194$) and \textit{Business and Career Advice} ($n=121$) were mentioned by 28.6\% and 17.85\% of LLM learners, respectively. \textit{Service Utilization} ($n=161$) was reported by 23.75\% LLM learners. Other inquiries were mentioned by less than 20\% of LLM learners, as per \autoref{fig:learning-inquiries}.

Following this, we zoomed into the learning scenario and inquired about the learning tasks participants complete with the assistance of LLMs, presenting them with a multiple-choice list of options. Participants could then report the frequency of use for each selected task on a scale from \textit{Less than once a month} to \textit{Every day}. 

Participants engage with LLMs for a wide range of \textit{learning} tasks. On average, participants report using LLMs for 7 different tasks ($M = 6.99$, $SD = 3.42$). However, the distribution is broad, with some individuals using LLMs for as few as one task, while others engage in up to 17 different tasks ($\tilde{x}=6$).

\begin{figure}[h]
    \centering
    \includegraphics[width=\linewidth]{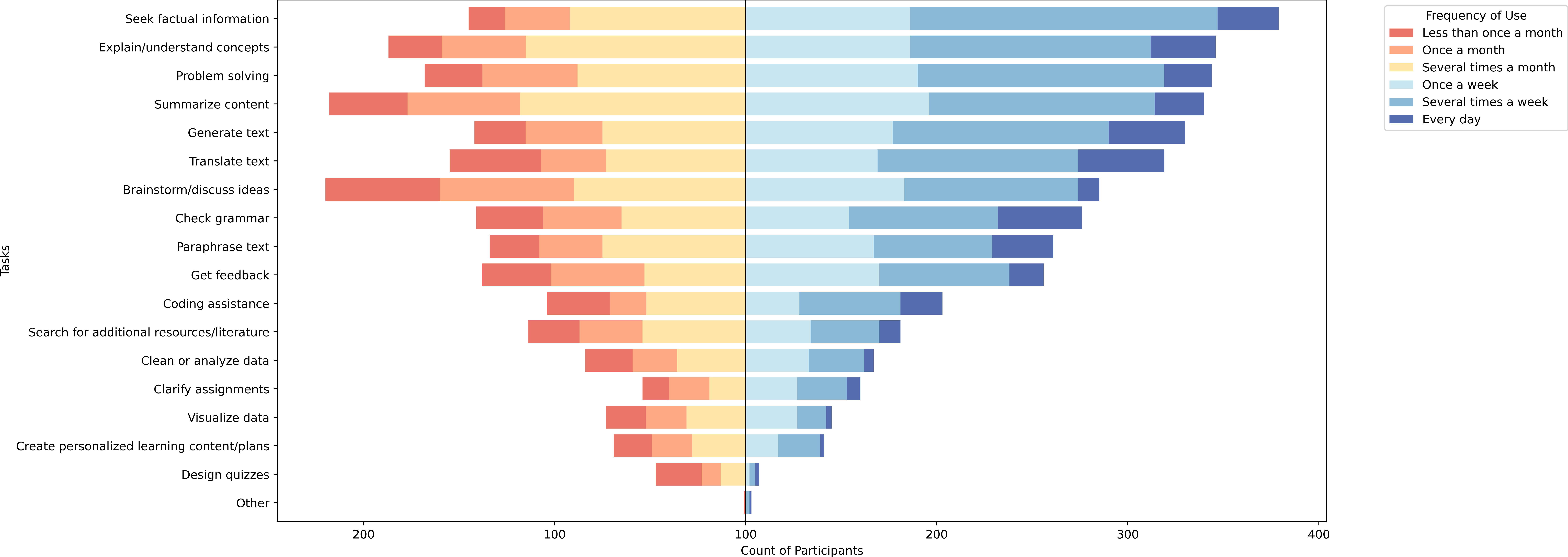}
    \caption{Frequency of the learning tasks participants perform with LLMs}
    \label{fig:learning-tasks}
\end{figure}

As presented in \autoref{fig:learning-tasks}, learning tasks related to \textit{content summarization, concept explanation}, and \textit{factual information retrieval} are among the most frequently performed, with many participants engaging in these activities \textit{at least weekly}.

Brainstorming, problem-solving, and text-related tasks (\textit{translation, paraphrasing,} and \textit{grammar checking}) also see frequent usage, though with greater variation: while some learners use LLMs regularly, others engage with these tasks only occasionally or monthly. More specialized tasks, such as \textit{coding assistance, data analysis,} and \textit{quiz design}, are performed less frequently, with most users engaging less than once a month or only on an as-needed basis. \textit{Clarifying assignments} and \textit{searching for additional resources} are also among the less frequent use cases.

\paragraph{Interaction Modalities} 

Today's LLMs are multimodal, accepting images or audio files besides text, too, so we asked which interaction modalities users employ. Virtually all learning-users (97.3\%) use \textit{text-to-text} interaction, that is, typing text questions and reading text answers. This is followed by using images as part of the LLM-supported learning process: either by asking the AI to interpret or describe images (i.e., \textit{image-to-text}, $n=193$) or by generating images from text prompts (i.e., \textit{text-to-image}, $n=145$). Some participants indicated using voice interfaces to converse with the LLM, either with voice input (\textit{speech-to-text}, $n=113$) or having the LLM answer with speech (\textit{text-to-speech}, $n=74$). Far fewer have ventured into \textit{video-to-text} ($n=29$) or \textit{text-to-video} ($n=21$) interaction.

\subsubsection{The When, Where, and How of Everyday Learning with LLMs}

\paragraph{Duration \& Frequency}
63.5\% of participants utilize LLMs for a duration exceeding six months: 34.4\% ($n=233$) engaging for more than a year and 29.2\% ($n=198$) for a period between six months and a year. In contrast, only 4.6\% ($n=31$) of participants reported utilizing LLMs for less than one month. The median self-reported time of use was 20 minutes per week spent interacting with LLMs for learning tasks ($M=47.2$ min/week, $SD=80.36$ with some heavy users spending many hours $max=900$). 20\% ($n=137$) of LLM learners stated using LLMs every day for learning for at least one task with LLMs (more on tasks in \autoref{sec:tasks}). Most participants however, almost 60\% ($n=406$), use LLMs at least once a week (but not daily). Less than 3\% ($n=18$) of participants use LLMs exclusively less than once a month. 

\begin{figure}[t]
    \centering
    \includegraphics[width=.75\linewidth]{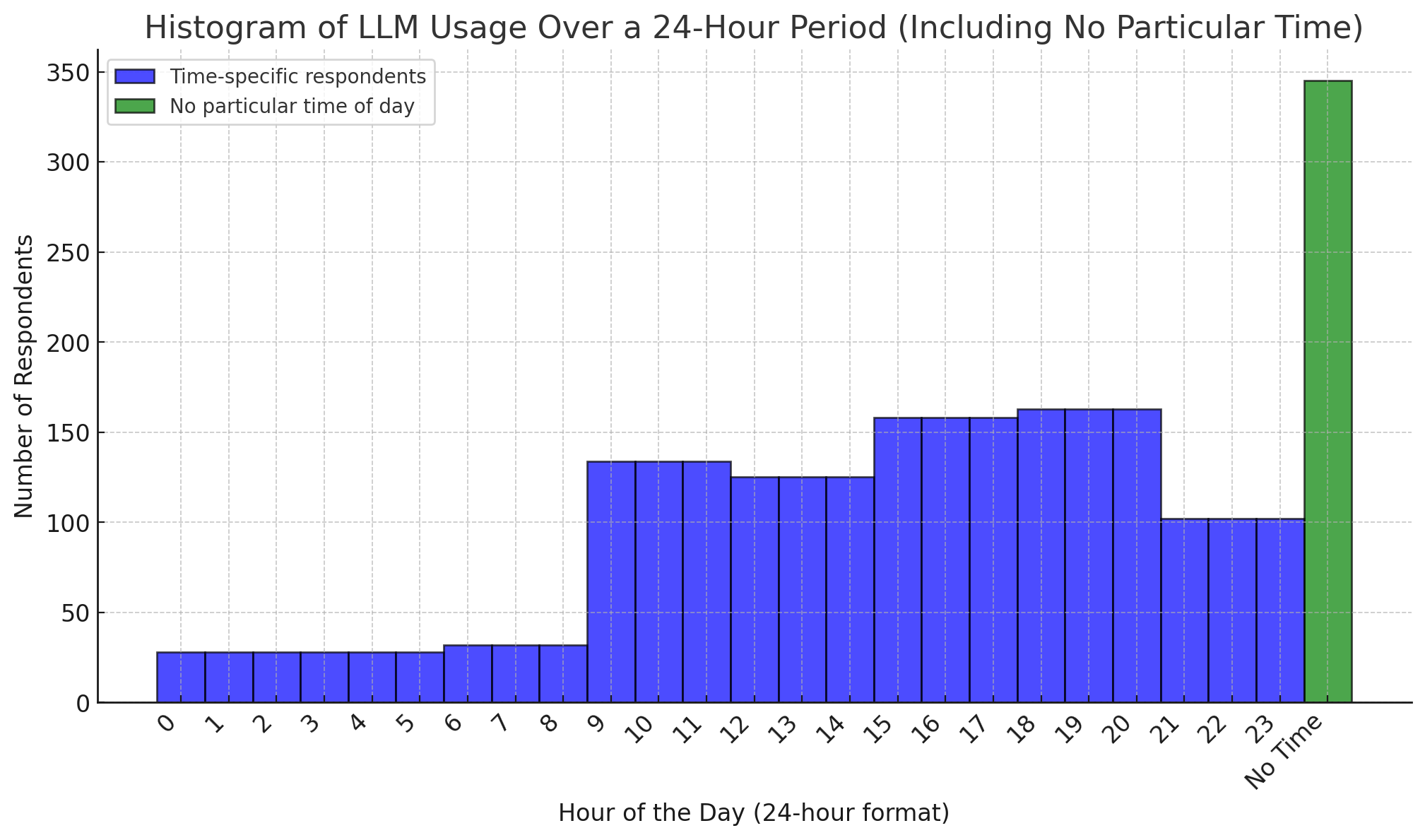}
    \caption{Participants' hourly distribution of LLM use for learning over a 24-hour period}
    \label{fig:histogram-day-distribution}
\end{figure}

\paragraph{Timing} Most LLM learners do not reserve a special time or day for LLM-based learning, that is, they use it whenever the need pops up. Nearly 70\% ($n=471$) selected \textit{no particular day} of the week for using LLMs, and about 51\% ($n=345$) \textit{no particular time of day.} 
Among those LLM learners who did specify timing, weekday evenings and afternoons were a bit more common (see \autoref{fig:histogram-day-distribution}): for instance, evening (6–9 p.m.) was the most cited specific time slot (by 24\% of respondents, $n=163$), followed by late afternoon (3–6 p.m., 23\%, $n=158$). Very few (under 5\% each) used LLMs primarily in the early morning ($n=32$) or overnight hours ($n=28$). As for days, weekdays slightly outranked weekends – each weekday was cited by between 19\% and 23\% of participants, whereas only 11\% mentioned Sundays.

In terms of \textit{how much time} learners engage with LLMs on an average weekday, over half of LLM users ($n=356$, approximately 53\%) reported spending less than 30 minutes per day using LLMs for learning. About 21\% ($n=141$) use them for 30–60 minutes daily, and roughly 15\% ($n=104$) for 1–2 hours per day, while 11\% ($n=77$) are power-users exceeding 2 hours a day.

\paragraph{Context}

LLM learners could select one or more learning contexts in which they engage with LLMs. The results show integrating these AI tools across a variety of learning contexts. On average, participants reported learning in 2.6 different contexts ($SD = 1.28$), with the number of contexts per person ranging from 1 to 6 ($\tilde{x}=2$). 25\% of participants selected 3 or more learning contexts. 

The variety of contexts is led by \textit{Higher Education} ($n=71$) and the combinations of \textit{Work-Related Activities--Practical Skills Learning-- Lifelong/Hobby Learning} ($n=35$), and \textit{Practical Skills Learning--Lifelong/Hobby Learning} ($n=32$). Across all responses, work and personal learning are both major domains: just over half of users (around 50 to 55\%) use LLMs for \textit{Lifelong/Hobby Learning} ($n=367$) – i.e., self-directed learning of personal interests – and a similarly high percentage use them for \textit{Work-Related Activities} ($n=365$), i.e., on the job for learning or problem-solving at work. Nearly half, 48\% ($n=325$), employ LLMs for \textit{Practical Skills Learning}, followed by \textit{Higher Education} with 47\% ($n=322$). 
Less commonly selected contexts included \textit{Professional/Career Development} ($n=231$), \textit{Popular Science Education} ($n=110$), \textit{K-12 Education} ($n=25$, probably due to our adults' sample) and \textit{Other} learning contexts, such as language learning or health inquiries, with one participant stating \quotes{introspective activities}.

\begin{figure*}[h]
\centering
    \begin{subfigure}[t]{0.33\textwidth}
        \centering
        \includegraphics[width=.95\linewidth]{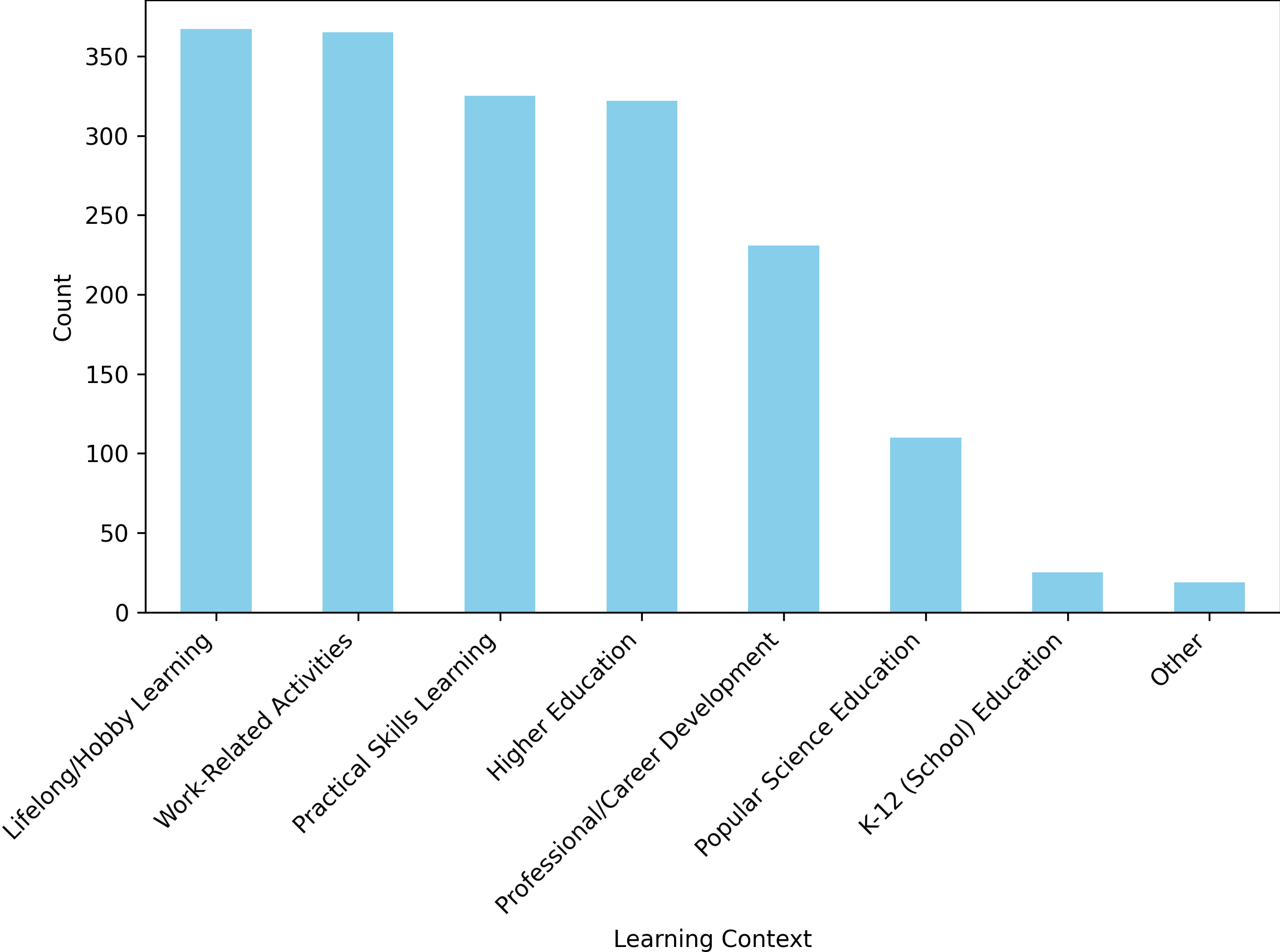}
        \caption{LLM-learning contexts}
        \label{fig:learning-contexts}
    \end{subfigure}
    \hfill
    \begin{subfigure}[t]{0.33\textwidth}
        \centering
        \includegraphics[width=.95\linewidth, trim = 0 -1.7cm 0 0]{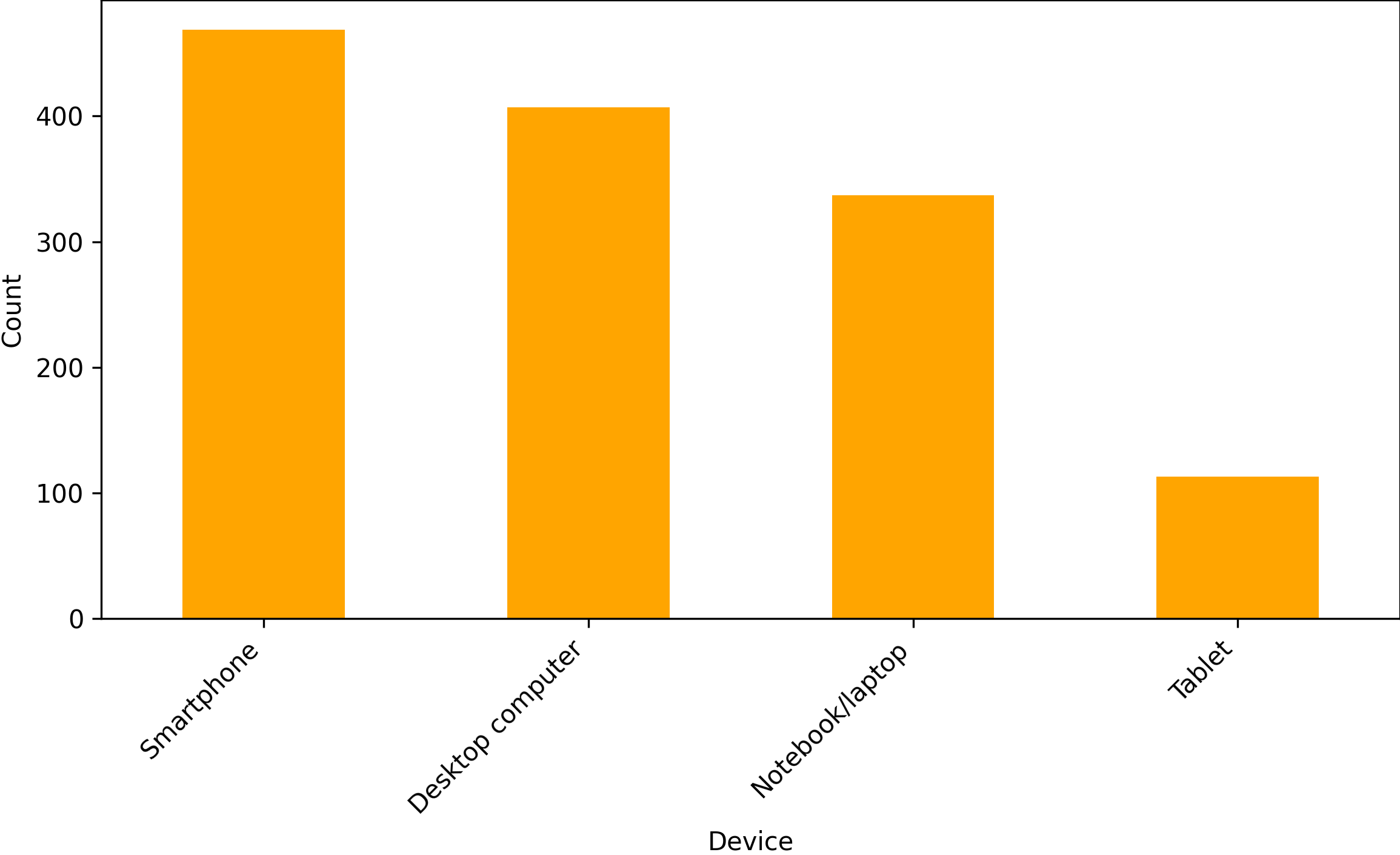}
        \caption{Devices to access LLMs}
        \label{fig:device}
    \end{subfigure}
       \hfill
    \begin{subfigure}[t]{0.33\textwidth}
        \centering
        \includegraphics[width=.95\linewidth]{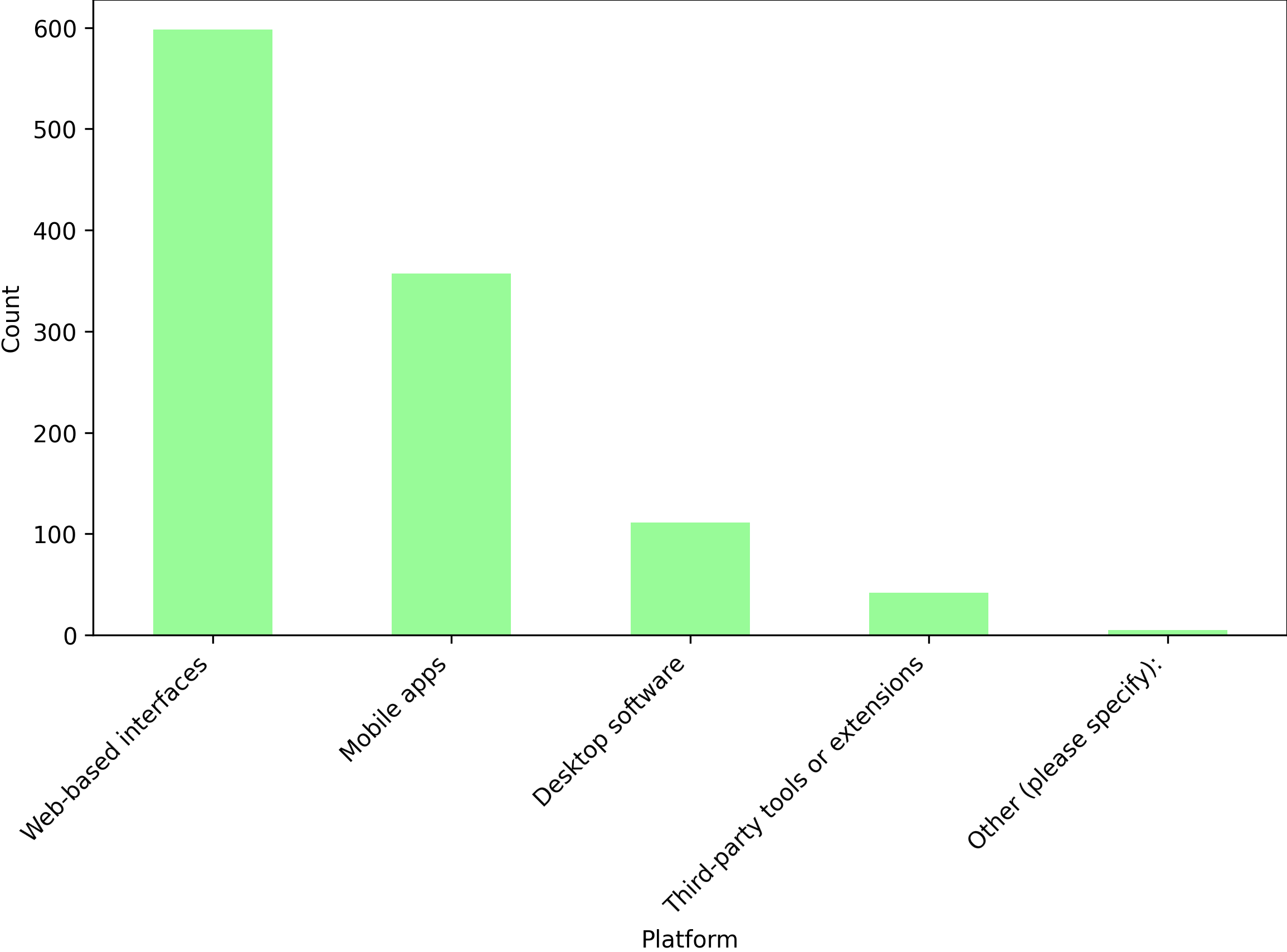}
        \caption{Platforms to access LLMs}
        \label{fig:platforms}
    \end{subfigure}
    \caption{Frequency of the mentioned a) learning contexts with LLMs  b) devices to access LLMs for learning and c) platforms to access LLMs for learning}
    \label{fig:contexts-devices-platforms}
\end{figure*}

\paragraph{Social Setting} 
LLM-based learning is almost entirely an individual activity. An overwhelming 89\% ($n=606$) of respondents said they primarily use LLMs for learning alone (individual learning). Only 1\% ($n=8$) primarily use them in a group or collaborative setting. Another 9\% ($n=64$) said both individual and group use in equal measure. 

\paragraph{Devices and Platforms}
Users access LLMs through two devices on average, with smartphones and computers dominating. Smartphones were the most commonly used device (about 69\% of learners use a smartphone for LLM access, $n=469$), closely followed by desktop PCs (60\%, $n=407$) and laptops (50\%, n=$337$). Tablets were used by a smaller share (17\%, $n=113$). 
As for how LLM learners access LLMs, the vast majority (about 88\%, $n=598$) use standard web-based interfaces. Over half (53\%, $n=357$) also reported using mobile apps (for instance, the official ChatGPT app or third-party client apps). Around 16\% have used desktop software integrations ($n=111$), and a small fraction use browser extensions or other third-party tools (6\%, $n=42$). As such, most participants are using the readily available free interfaces, with web and mobile apps being the go-to methods of access. 

\subsubsection{Latent Learner Profiles}

We ran a LCA that resulted in four profiles of LLM learners based on their learning behaviors, task engagement, and digital environments. The optimal four‐class solution was selected based on the log‐likelihood (maximum likelihood = –10575.98), AIC (21397.96), and BIC (21953.82). Estimated class population shares were relatively balanced (23.8\%–25.7\%), and modal posterior probabilities confirmed the robustness of the solution.

The first group, \textit{Structured Knowledge Builders} (23.8\%), primarily engages with LLMs in higher education contexts, with 71\% using them for academic learning. Their LLM usage is focused on conceptual understanding (67\%), text generation (44\%), and summarizing content (77\%), indicating that they use AI tools to synthesize and deepen their academic knowledge. While they occasionally engage in work-related activities (47\%), they are less likely to rely on LLMs for tasks such as coding or assignment clarification. This group predominantly uses laptops (80\%) but also incorporates smartphones (53\%) into their learning.

In contrast, \textit{Self-Guided Explorers} (26.1\%) demonstrate a strong preference for informal, self-directed learning. They are the most active in lifelong learning and hobby-related education (85\%) and show high engagement in practical skills learning (65\%), whereas their use of LLMs in formal education is limited. Their primary focus is on problem-solving (54\%), seeking factual information (72\%), and developing practical competencies rather than structured academic tasks. With an overwhelming reliance on smartphones (91\%), they represent a highly mobile and on-the-go learning style, leveraging LLMs as an everyday knowledge companion.

The third profile, \textit{Analytical Problem Solvers} (24.4\%), is characterized by a stationary, office-based learning environment, with nearly all members (98\%) using desktop computers. They engage moderately in both higher education (45\%) and work-related activities (47\%), indicating a mixed academic and professional focus. While their overall engagement with LLMs is lower than that of other groups, they rely on AI for problem-solving (64\%) more than for creative or generative tasks. Their limited engagement in brainstorming, grammar checking, or text summarization suggests that they use LLMs more for analytical reasoning and structured problem-solving rather than for open-ended content creation.

The fourth and most versatile group, \textit{Adaptive Power Users} (25.7\%), demonstrates a highly integrative approach to LLM usage. These learners are present in multiple contexts, with increased engagement in higher education (59\%), lifelong learning (69\%), professional development (55\%), and work-related activities (68\%). Unlike the other groups, they exhibit consistently high engagement across all major learning tasks: 84\% use LLMs for brainstorming and discussion, 78\% for grammar checking and text generation, 87\% for explaining concepts, and 94\% for summarizing content. Their diverse learning behaviors are reflected in their device usage, where they incorporate smartphones (81\%), desktops (65\%), and laptops (59\%), indicating a flexible and adaptive approach to AI-assisted learning across different settings.

\begin{figure}[t]
    \centering
    \includegraphics[width=\linewidth]{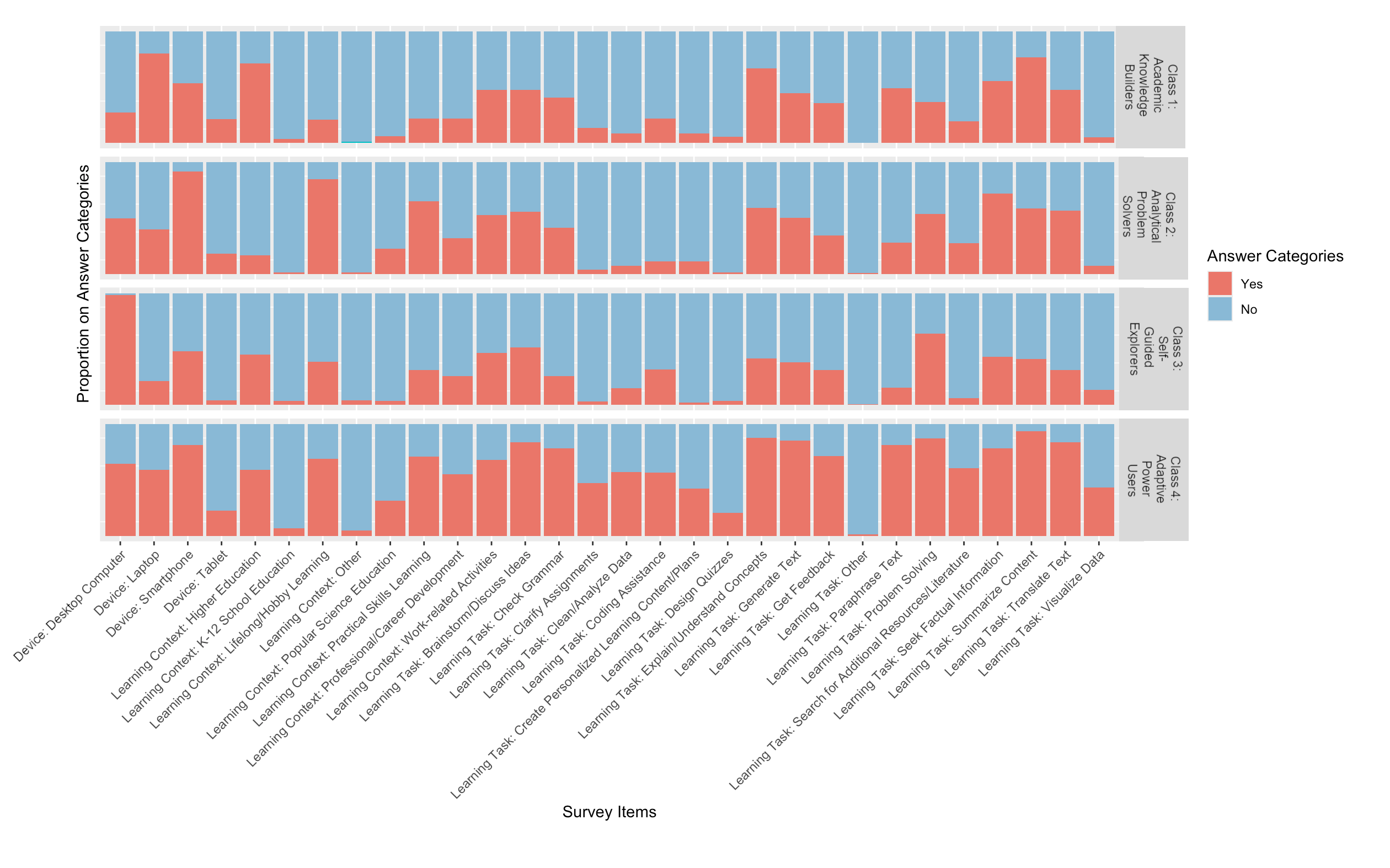}
    \caption{Latent Class Analysis: Survey Item Probabilities by Learner Profile: Grouped by Profile}
    \label{fig:LCA4}
\end{figure}

\begin{figure}[t]
    \centering
    \includegraphics[width=\linewidth]{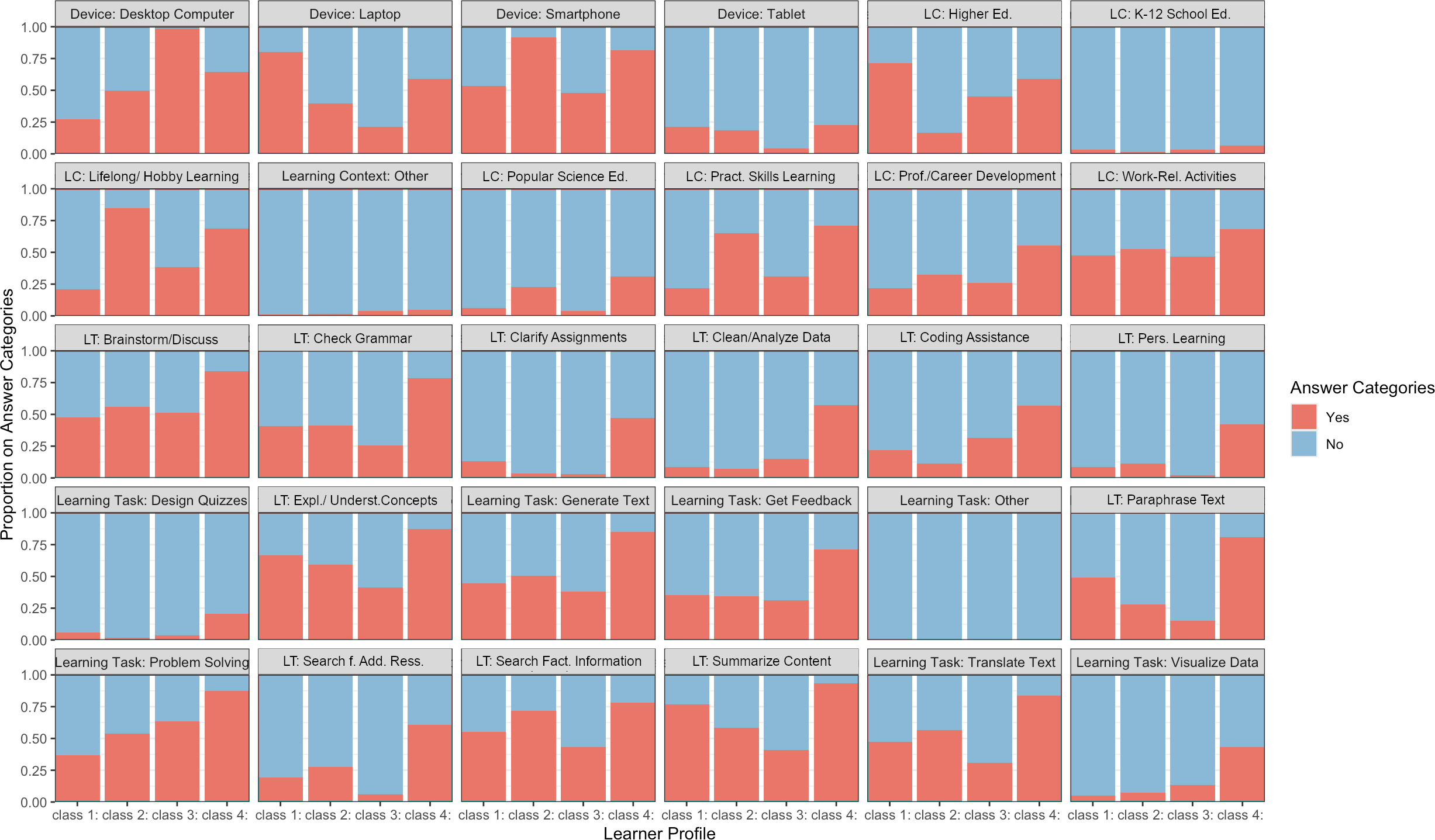}
    \caption{Latent Class Analysis: Survey Item Probabilities by Learner Profile: Grouped by Survey Item}
    \label{fig:LCAvar4}
\end{figure}

Whereas \textit{Academic Knowledge Builders} and \textit{Analytical Problem Solvers} tend to operate within more defined academic and professional contexts, \textit{Self-Guided Explorers} embrace informal learning with a mobile-first approach, and \textit{Adaptive Power Users} integrate AI tools into a wide range of learning contexts. 
Regarding the tasks the learners execute with LLMs, \textit{Structured Knowledge Builders} focus on content generation and conceptual learning within academic settings, but have lower engagement in problem-solving and technical applications. \textit{Self-Guided Explorers} prioritize on-demand, self-directed learning, using LLMs for fact-checking, problem-solving, and resource exploration, but rarely for structured assignments or academic writing. \textit{Analytical Problem Solvers} employ LLMs for technical and computational tasks, with less emphasis on brainstorming, writing, or general conceptual explanations. \textit{Adaptive Power Users}, in contrast, are highly versatile and engage in nearly all task types, making them the most comprehensive users of LLMs. 

\summarize{\paragraph{\textbf{RQ3 Main Findings}}
The timing and context data, next to the LCA, suggest that LLMs are empowering \textbf{self-directed, just-in-time and life-centric learning} opportunities. Most usage is solitary and on personal devices like smartphones and laptops, using existing platforms and interfaces to access LLMs. The fact that over half of users have no fixed time or schedule for it might point to LLMs filling micro-learning moments throughout day-to-day life. The dominant interaction modality remains the initial text-based manner.

LLM learners are leveraging LLMs to save time and get unstuck. Summarizing complex content, explaining difficult concepts in simpler terms, providing quick factual answers, and brainstorming ideas are among the top uses. Here, we observe a curious contradiction: the concern of misinformation seems \textbf{not to hinder our participants in utilizing LLMs for factual checks.} 

Among LLM learners, we identified four classes: \textit{Structured Knowledge Builders} use LLMs mainly for academic tasks like summarization and conceptual understanding, \textit{Self-Guided Explorers} rely on mobile devices for informal, practical learning, and \textit{Analytical Problem Solvers} apply LLMs to technical problem-solving in stationary, work-based contexts. In contrast, \textit{Adaptive Power Users} engage across all learning contexts and task types, showing the most comprehensive and flexible use of LLMs. }

\subsection{RQ4: Perceived Learning Effects, Over-reliance and Privacy Measures}

\subsubsection{Perceived Learning Effects}
\label{sec:productivity}
 
\begin{figure}[t]
    \centering
    \includegraphics[width=\linewidth]{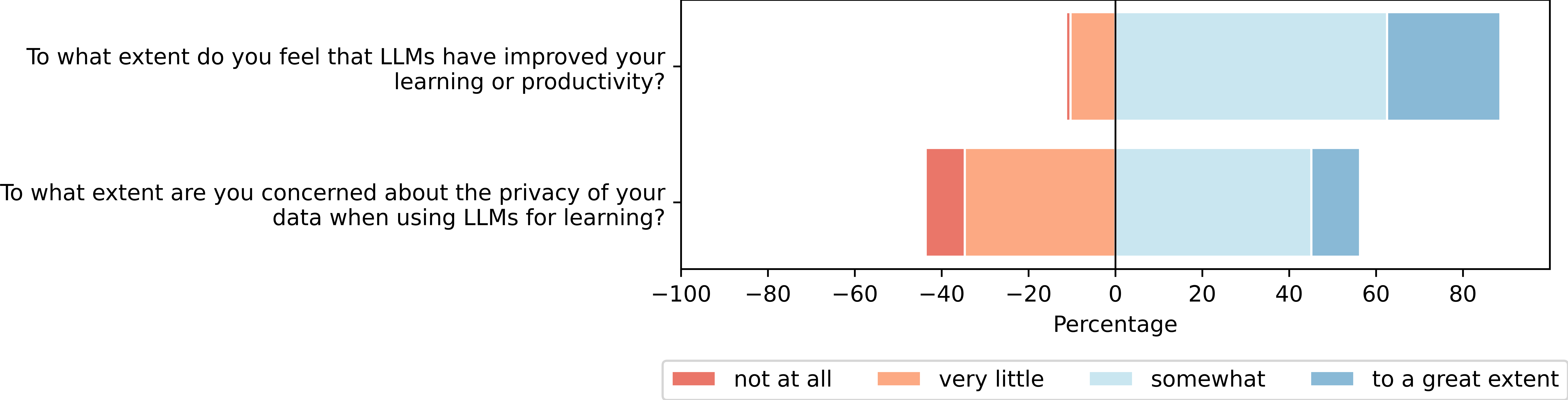}
    \caption{Likert-scale distribution of LLM perceptions on productivity (above) and privacy concerns (below)}
    \label{fig:effectiveness-privacy}
\end{figure}

The surveyed LLM learners shared their opinion on the extent LLMs have improved their learning and productivity on a 4-point Likert-scale, from \textit{to a great extent} to \textit{not at all}. A short open-text question probed participants to explain their answer briefly. Our results indicate that LLM learners hold nuanced attitudes about using these tools for learning. They generally recognize LLM's benefits, whilst being somewhat mindful of risks such as privacy and dependency, as depicted in the upper barchart in \autoref{fig:effectiveness-privacy}. 

88\% ($n=601$)  of participants perceive LLMs to have improved their learning or productivity, with 177 participants agreeing it to be to a great extent ($n=424$ replying \textit{somewhat}). Participants emphasized speed, convenience, instant access, more information, new ideas, and easy understanding to be the major drivers of improved learning processes and increased productivity. 

\quotes{LLMs [help] me to find many new ideas and [give] me a lot of information that I didn't know before. Its function is really amazing and always available whenever I need it.} (P140) 

\quotes{LLMs allowed me to navigate content in a more efficiency and simpler manner. This reduced significantly the frustration in the learning process in some contexts - for example: being stuck understanding a certain concept because it is explained in a complicated manner.} (P28)

\quotes{Spending time reading a book, through papers or googling around to find a semi-good answer takes so much more time than using an LLM.} (P442)

\begin{figure}[t]
    \centering
    \includegraphics[width=\linewidth]{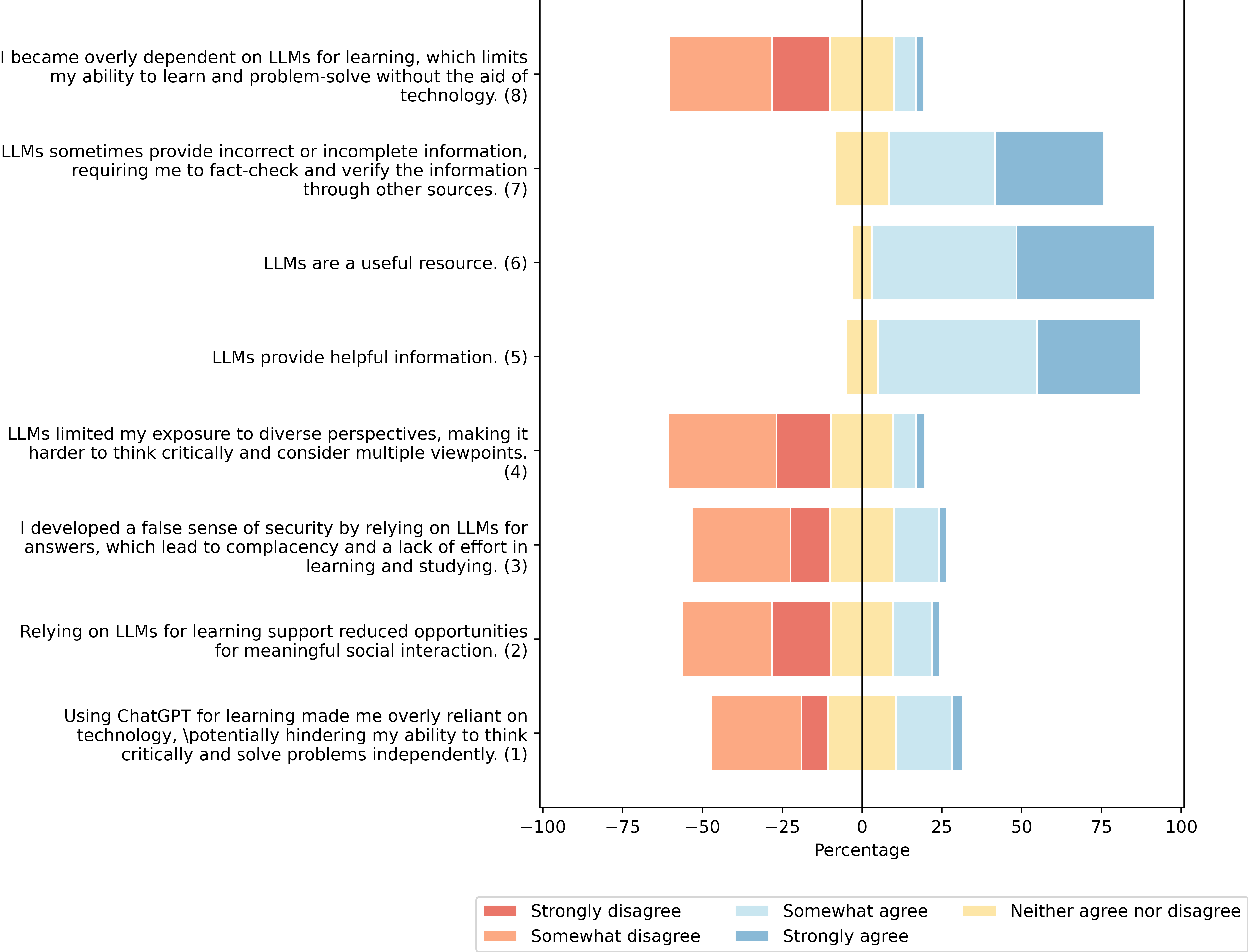}
    \caption{Likert-scale distribution of LLM perceptions of over-reliance and usefulness}
    \label{fig:perception-overreliance}
\end{figure}

\subsubsection{(Over-)Reliance on LLMs}
We evaluated eight Likert-scale statements, from strongly disagree to strongly agree, related to the over-reliance on LLMs for learning, including concerns about misinformation, usefulness, and social impact. The descriptive results are presented in \autoref{fig:perception-overreliance}. 

Responses regarding over-reliance on LLMs and reduced critical thinking are rather on the pro-LLM side, with a notable proportion disagreeing that LLMs hinder independent problem-solving. Similarly, concerns about diminished social interaction due to LLM use elicit mixed responses. The majority of participants agree that LLMs provide helpful and informative content, yet they also acknowledge that misinformation remains a challenge, requiring fact-checking. These findings overlap with the emerged challenges listed in \autoref{sec:challenges} above. 

\subsubsection{Privacy Attitudes}

We asked participants to what extent they have privacy concerns when interacting with LLMs on a 4-point Likert-scale from \textit{not at all} to \textit{to a great extent}. 

We observe a low to moderate level of privacy concern, with a \textit{somewhat} degree of concern ($n=306$), followed by \textit{very little} ($n=235$) being the majority (see \autoref{fig:effectiveness-privacy}). Very few users are on the opposite sides of concerns, with similar numbers on both: 76 participants express a concern \textit{to a great extent}, whereas 61 participants bear \textit{no concern at all}. 

When asking whether participants incorporate specific privacy measures, the majority ($n=543$) indicated that they do not take such measures, compared to 135 participants who confirmed implementing specific privacy measures.


\subsubsection{Continuance of LLM Use for Learning}

\begin{figure*}[t]
\centering
    \begin{subfigure}[t]{0.62\textwidth}
        \centering
        \includegraphics[width=.95\linewidth]{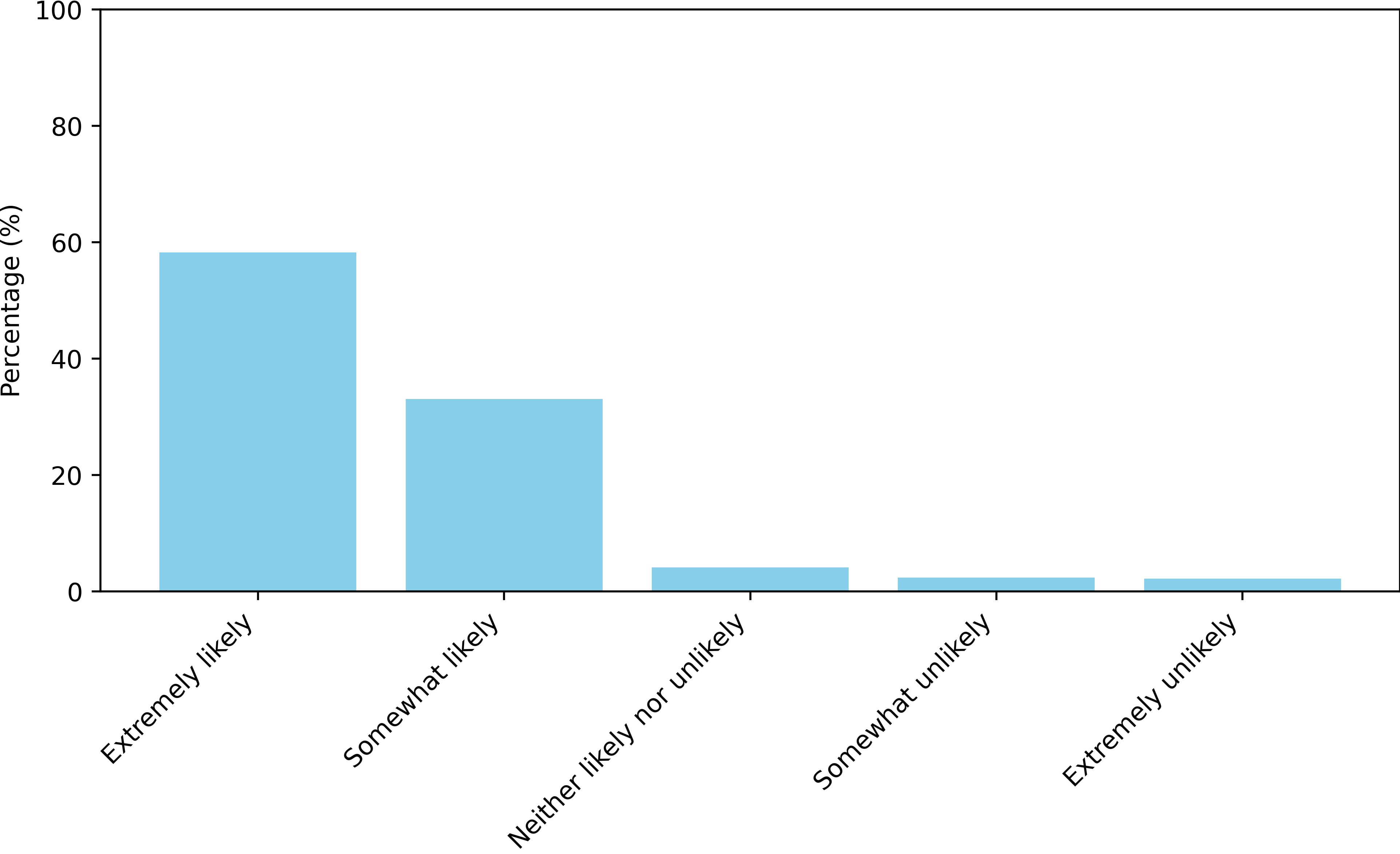}
        \caption{Likeliness of continuance of using LLMs for learning}
        \label{fig:likeliness-continuance}
    \end{subfigure}
    \hfill
    \begin{subfigure}[t]{0.37\textwidth}
        \centering
        \includegraphics[width=.95\linewidth, trim=0 -2.5cm 0 0]{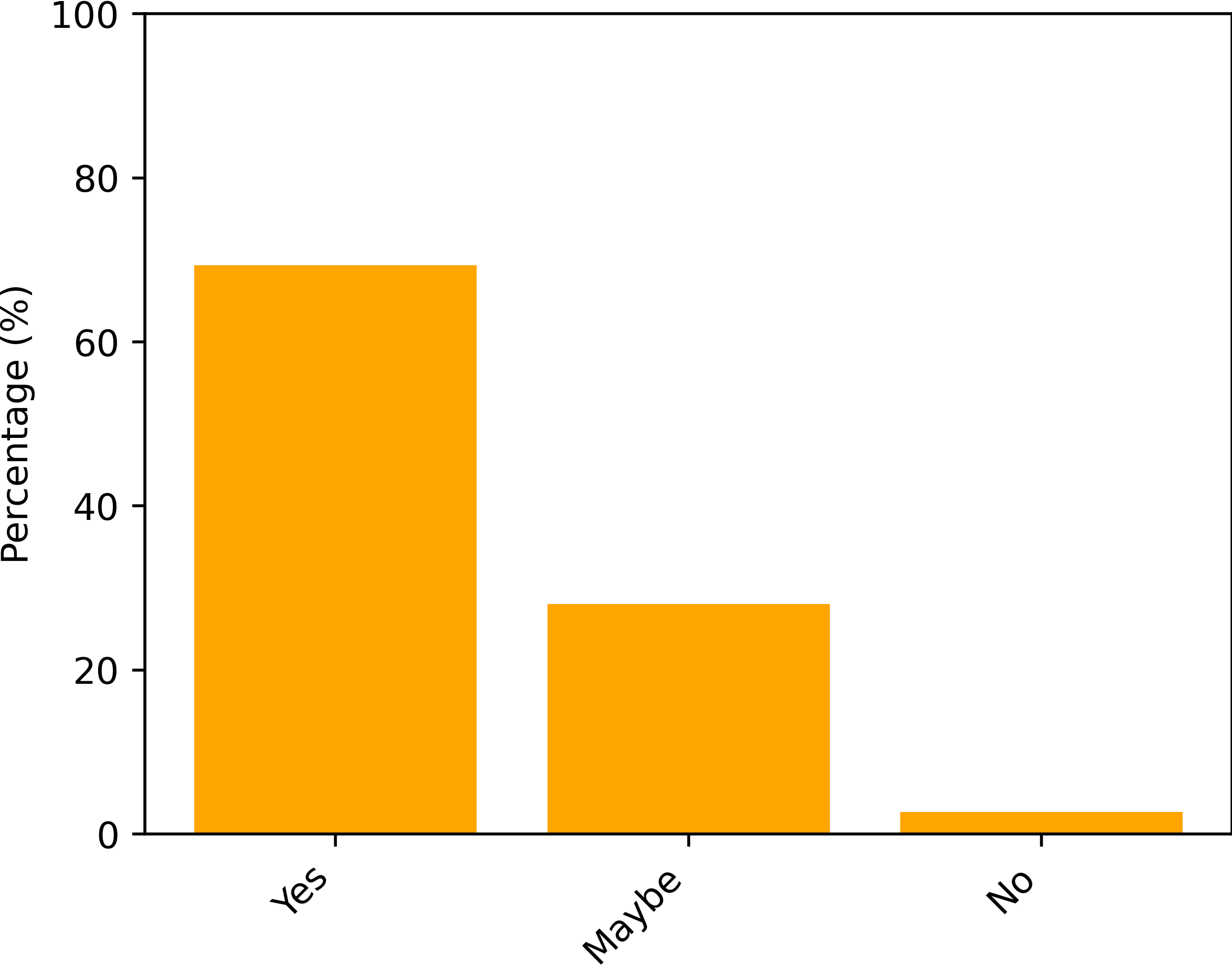}
        \caption{Recommendations of LLM for learning}
        \label{fig:recommendation-llms-learning}
    \end{subfigure}
    \caption{LLM learners' responses on a) how likely they are to continue using LLMs for learning and b) whether they would recommend using LLMs to friends}
    \label{fig:future-use}
\end{figure*}

Finally, we asked participants to state on a 5-point Likert scale the likelihood of them continuing to use LLMs for learning. In line with previous positive responses, a large majority of 58\% ($n=395$) consider themselves \textit{extremely likely} to continue using LLMs for learning, followed by one-third ($n=224$) who say they are \textit{somewhat likely}. Far fewer respondents, 4\%, ($n=28$) are neutral, and less than 5\% combined are somewhat unlikely ($n=16$) or extremely unlikely ($n=15$), as \autoref{fig:likeliness-continuance} depicts. In a similar vein, 72\% would recommend LLMs for learning to their friends (see \autoref{fig:recommendation-llms-learning}).

\subsubsection{Regression Analysis}

We ran a multinomial logistic regression to assess how perceptions of privacy, effectiveness, overreliance, and demographic factors (age, gender, AI literacy, ATI score) predict latent class membership. Class \textit{Adaptive Power Users} served as the reference category.
The model was statistically significant overall ($\chi2(21) = 156.4, p < .001$) 
with a pseudo R² of 0.083, indicating modest explanatory power.
Perceived effectiveness of LLMs emerged as a robust and consistent predictor across comparisons. Higher effectiveness ratings significantly decreased the likelihood of membership in class \textit{Structured Knowledge Builders} ($\beta = -1.11, p < .001$), class \textit{Self-Guided Explorers} ($\beta = -0.72, p < .001$), and class \textit{Analytical Problem Solvers} ($\beta = -1.06, p < .001$) compared to class of \textit{Adaptive Power Users}. Similarly, more favorable attitudes toward technology (ATI score) predicted higher odds of membership in the Class of \textit{Adaptive Power Users} over all other classes (p < .01 across comparisons).
Overreliance concern negatively predicted membership to class Self-Guided Explorers ($\beta = -0.24, p = .015$), while age was positively associated with the same class($\beta = +0.038, p = .001$) and marginally negatively associated with Class \textit{Structured Knowledge Builders}(p = .095).

We extended our multinomial logistic regression to include participants' payment status for LLM tools, in addition to their attitudes and demographic factors. The model remained significant ($\chi2(24) = 177.2, p < .001$) and explained nearly 10\% of the variance in latent class membership (pseudo R² = .094).
Perceived effectiveness continued to strongly predict membership in more engaged classes, with higher ratings significantly reducing the likelihood of belonging to classes other than the comparative Class \textit{Adaptive Power Users}. Technological affinity (ATI Score) and payment behavior also emerged as consistent and meaningful predictors. In particular, participants who paid for LLMs were significantly more likely to belong to \textit{Adaptive Power Users} and less likely to be in the lower-engagement classes. Gender significantly predicted membership to the class \textit{Analytical Problem Solvers}, with male participants nearly three times as likely to belong to that group ($\beta = 1.07, p < .001$).

\summarize{\paragraph{\textbf{RQ4 Main Findings}}
Most LLM learners perceive LLMs as helpful and useful resources for improving learning and productivity. Interestingly, LLM learners mostly disagree with the over-reliance narrative of LLMs impeding critical thinking and meaningful social interaction. Our pool of LLM learners exhibits no privacy concerns to a great extent. It thus comes as no surprise that the large majority indicates a strong willingness to continue using LLMs for learning, suggesting that, into their third year of roll-out, LLMs might have become the \textit{new normal} for learning.  }

\section{Discussion}

\subsection{Widespread and Embedded Use of LLMs in Everyday Learning -- with Technical Affinity as the Main Driver of Adoption}

Our survey reveals a prominent observation: 88\% of our respondents reported using LLMs as part of their learning process, be it for professional development or as voluntary academic support. This high adoption rate shows a clear shift nearly three years after the widespread distribution of LLMs in late 2022: LLMs are no longer emerging technologies, but rather deeply embedded tools in people's everyday learning routines. 

Yet, this broad uptake is not evenly distributed across different demographic groups in our sample. Our findings reveal significant differences between LLM learners and non-LLM learners regarding age, gender and technical affinity attitudes, suggesting that LLM learners tend to be younger, male and more comfortable with novel technology, which aligns with our and similar findings \cite{strzelecki2024use} of curiosity and openness being the main driver of initial LLM adoption. However, our more nuanced latent class analysis of LLM learners shows a rather equal gender distribution across the classes of \textit{Self-guided Explorers} and \textit{Analytical Power Users}. In other words, once individuals begin using LLMs for learning, the way and depth they engage with them is less strongly shaped by gender. This highlights differences between adoption and engagement: while curiosity and tech confidence may influence who starts using LLMs, other factors might shape how those tools are ultimately used for learning. For example, \citet{draxler2023gender} found the gender gap in LLM use to be of less size in technical fields,  suggesting that proper education on LLM use itself might be a strong potential avenue to tackle the digital divide. Similarly, awareness of the benefits, but also limitations and risks of LLMs all positively correlate with the intention to use LLMs. In other words, users possessing the knowledge on what LLMs can do wrong does not necessarily hinder their use of LLMs\cite{almurshidi2024understanding}.  
Indeed, our results show that 14\% of non‑adopters ($n=14$, combined non-LLM users and non-LLM learners) within our participants' pool simply do not know how to apply LLMs to their learning, which inhibits initial adoption -- despite non-significant differences in knowledge regarding the operating mechanisms of AI (i.e., AI literacy results). This points to a gap not in theoretical knowledge, but in practical, confidence-building exposure. To overcome this barrier, we need both real‑world demonstrations of LLMs' educational value, as well as AI literacy programs that prioritize practical, everyday learning benefits over technical AI capability, that are developed and approved by proven educational authorities. 

\subsection{Convenience over Accuracy and Privacy}

Our participants identified mistrust and disinformation as the primary challenges when using LLMs for learning. LLMs are indeed prone to producing hallucinations \cite{huang2025hallucination}, which are errors in generated content. These can be categorized into factual hallucinations, where the content deviates from real-world facts, and faithfulness hallucinations, where the generated content does not align with user instructions or contains internal inconsistencies \cite{huang2025hallucination}. Hallucinations in LLMs present a critical challenge for learning, as errors in generated content can often be difficult to detect, especially for novice users \cite{bender2021parrots}. Particularly over time, there is a risk of diminished critical engagement with content and an accumulation of misconceptions. Despite this awareness, though, our findings reveal \textit{fact-checking} to be the most common reported LLM use case: 301 learners in our sample suggest using LLMs to verify information while simultaneously expressing concerns about misinformation. 
This contradiction may indicate users swapping accuracy for convenience and speed in learning, as our results demonstrate an around-the-clock, ubiquitous use of LLMs for learning.
 
LLMs can be \textit{convincingly wrong} \cite{skjuve2023uxchatgpt}, and people are more likely to trust AI if the generated content matches their own worldviews \cite{shahid2022fakevideos}. As LLMs begin to integrate internal fact-checking tools, such as Google's DataGemma feature\footnote{\url{https://ai.google.dev/gemma/docs/datagemma}} that allows models to verify their own responses, users may place even more trust in these systems. However, fact-checking is not exclusive to LLMs. The widespread distribution of social media has long enabled the spread of fake news \cite{flintham2018fakenews}. In a study examining the consumption and trustworthiness of news on social media, \citet{flintham2018fakenews} found their participants to rely on objective factors (e.g., source URL, journalistic style) and subjective factors (e.g., personal interests in the topic) when judging news as fake or not. For LLMs, the first might be more difficult to follow, given that LLMs do not name their sources and do not operate like that in the first place. The lack of transparent sourcing can further impede users' ability to verify information, increasing the risk of internalizing inaccuracies. Future research should thus explore how learners recognize hallucinations and evaluate credibility, what strategies they use to cross-check information, and how cognitive biases shape their trust in AI-generated content.

Another barrier to using LLMs is privacy, as evidenced by participants who explicitly cited data handling concerns. Despite these concerns, though, LLM learners reported low to moderate privacy concerns, with most of them not taking specific protective measures. Clearly, there exists a gap between users' attitudes and their actual behavior, again potentially suggesting the convenience of LLMs outweighs the concern over privacy. As an implication, LLM users might benefit from interfaces that encourage more privacy-conscious use, by, e.g., detecting, highlighting, or even anonymizing sensitive information in users' prompts. 

\subsection{Disparity in Over-reliance Perceptions}
Regarding learning effectiveness and overreliance on LLMs, \citet{lepp2025programming} surveys the frequency and methods of LLM use for learning programming skills among 231 computer science students. Their findings surprisingly reveal that more frequent reliance on AI chatbots for programming tasks is associated with a decline in students' performance. In a similar vein, \citet{STADLER2024cognitive} discovers LLMs to ease cognitive load, however, at the cost of depth in engaging with learning material, compared to the use of traditional search engines.
However, our findings complicate the narrative around overreliance. Despite the functional support that LLMs offer, participants in our survey did not express a strong sense of dependency or overreliance. Whereas the listed related works showcase that (excessive) use may correlate with performance decline, users themselves may not perceive their reliance as problematic, as our results suggest. This raises questions about the cause of this disparity. One possible explanation could lie in self-determination theory \cite{deci2012self}, that claims the importance of autonomy in motivation. Users may feel in control of their learning process even when the tool exerts a subtle influence, leading to a mismatch between perceived and actual reliance. Similarly, the Dunning-Kruger effect \cite{kruger1999unskilled} may play a role, where users with lower levels of expertise overestimate their ability to effectively use such tools without unintended consequences. One possible explanation for this reliance could be cognitive biases, such as the third-person effect \cite{davison1983third}, as a recent study on the LLM use in research writing suggests \cite{liao2024llmsresearchtoolslarge}. 

\subsection{Design Implications}

\subsubsection{Re-think Defaults Modes of Interaction for Learning}

In spite of the multimodal interaction possibilities, text-based interaction, known as 'prompting', remains the dominant modality of interaction with LLMs, as suggested by our findings. This comes as no surprise, given that LLMs operate on a natural language interface basis. As such, our results confirm previous findings that users tend to stick to default interface elements and interaction forms even when more graphic alternatives are made available \cite{dinner2011partitioning}. 
However, other media formats, when used for learning (e.g., visualizations or audio-based dialogue), bear immense potential to enhance the learning process \cite{lau2024wrapped, Mayer2009multimedialearning, kozma1991learning}. Rethinking the user interface of LLMs specifically for learning, to include multimodal learning materials by default, could potentially lead to a more in-depth analysis, understanding, and learning engagement. 

\subsubsection{Provide Learning Materials with Credible Sources and/or Confidence Levels}
As discussed above, our participants raised concerns about the tendency of LLMs to hallucinate, yet many did not limit LLMs' use for fact-checking. This reliance calls for transparent source attribution by, e.g., displaying source links, indicating confidence levels, or citing external references. Such features could support the users' agency by promoting awareness that AI can make mistakes and encouraging critical thinking skills for learners to question, verify, and reflect on the consumed media content.

\subsubsection{Cater to Specific Types of Learning and Learners} 
Current LLMs interfaces for learning (and in general) follow a one-size-fits-all approach for learning (and in general). However, our LCA uncovered four primary learner groups with LLMs, embodying different contexts of use, tasks, and devices of engagement. 
The diverse profiles underscore the contextual nature of LLM learning. This raises the question of whether adaptive interfaces could detect learner archetypes and tailor experiences accordingly. Although personalisation has the potential to support individual needs, there is also a risk that it could reinforce existing learning patterns and limit opportunities for growth. Rather than simply mirroring users' preferences, should adaptive systems gently challenge them? For example, should LLM interfaces gently nudge \textit{Self-Guided Explorers} toward more structured reflection, or support \textit{Analytical Problem Solvers} in developing creative fluency? 
Moreover, the mobility of learning, especially among \textit{Self-Guided Explorers}, highlights a shift toward ubiquitous, context-independent engagement with LLMs. This calls for mobile-first UI/UX optimization that potentially includes multimodal input, micro-interactions, or session continuity \cite{draxler2023thesis}. These can support learning in fragmented or informal settings, while addressing emerging needs like distraction management and coherence across short learning bursts. The \textit{Adaptive Power Users}, meanwhile, challenge one-size-fits-all design norms. As edge cases or design leaders, they reveal what is possible when AI tools are integrated flexibly and fluently. They may serve as testbeds for advanced features such as cross-tool integrations, longitudinal tracking, or scaffolding for complex reasoning, suggesting that we design not for the average, but with regards to advancements across a progression trajectory.

Finally, this discussion encourages us to consider equity and outcomes. Are all learners equally positioned to become \textit{Adaptive Power Users}, or do disparities in access, confidence, or prompt literacy widen existing gaps? And how do different usage patterns translate into actual learning gains? Does summarization encourage surface learning, while brainstorming supports deeper cognitive engagement? Such questions raise both ethical and motivational considerations around agency and autonomy in LLM use. 

\subsubsection{Empowering Learners to Become Prompt Engineers}

One downside of personalization and customization features in LLMs for specific learner profiles, which can initially support novice users, is that they may inadvertently constrain learners' development over time. Specifically, learners who rely heavily on automated adaptations may fail to develop transferable prompting skills, thereby limiting their ability to generalize their use of LLMs across contexts or evolve their interaction strategies as their expertise grows.

Our findings indicate that technical affinity, not basic AI knowledge, was a stronger predictor of learners' adoption of LLMs for learning. This suggests that the curiosity and willingness to experiment with new technologies, including the iterative refinement of prompts, are critical factors in facilitating meaningful and satisfying learning experiences. Importantly, this reinforces prior work showing that the quality of learner-generated prompts significantly affects the relevance and instructional value of LLM responses \cite{lyu2024evaluating, knoth2024ailiteracy}. This finding is further supported by a recent systematic review, which identifies prompt engineering as an emergent, teachable digital skill that improves student engagement and AI output quality in educational contexts \cite{lee2025prompt}, underscoring that prompting is not a peripheral activity but a central determinant of learning outcomes.

Rather than treating prompting as a side activity, interaction interfaces could embed reflective prompting practices, real-time feedback on input quality, and examples that illustrate the effects of prompt refinement within their interfaces. Various prompt engineering frameworks, such as CRISPE \cite{wang2024unleash}, offer structured approaches that can be embedded into learning interfaces to guide learners in formulating more effective interactions with LLMs. This approach, next to enhancing learners' agency, also aligns with broader goals of AI literacy and responsible tool use. As current practices suggest, empowering learners to become effective prompt engineers might become essential for realizing the long-term educational value of LLMs.

\subsubsection{Support Collaborative Learning by Design}

Our findings indicate that learning with LLMs is a predominantly individual activity, with only a small share of use dedicated to group learning. Although we did not examine the details of neither individual nor group learning sessions with LLMs, the large share already presents a limitation given the well-established benefits of group learning. Encouraging learning mechanisms that stem from group learning, such as pooled knowledge, reduced memory load, or observational learning \cite{nokes2015better}, are thus not being leveraged at the expense of benefits such as improved comprehension, retention, and problem solving \cite{nokes2015better}. 

Current mainstream LLM interfaces enable collaborative use only by sharing one account. These could be enhanced in several ways, with some proposals already mentioned in related work. For example, \citet{lyu2024evaluating} suggest collaborative query building, to construct and refine prompts in groups. These could be further enriched with response annotations, along with visible traces of who made the annotations. The system itself could integrate system- and context-aware prompts to encourage collaborative exchange (e.g., by facilitating a mediator role in learning exchanges or by making suggestions to form discussion groups out of currently active members). 

\subsection{Limitations}
Our sample, recruited exclusively from Germany via Prolific, may overrepresent tech-savvy, English-fluent individuals (mean AI literacy score = 3.2, line 573), thus limiting generalizability to non-Western or less tech-exposed populations. Cultural attitudes toward AI vary globally \cite{ge2024culture}, necessitating future multi-country studies. Additionally, Prolific participants' familiarity with research may introduce social desirability bias, though anonymity measures mitigated this. The use of ChatGPT for initial questionnaire design was validated through expert review and pilot testing, but future studies could employ cognitive interviews to further ensure question clarity and relevance.
In addition, in our future work, we will expand our survey to a larger geographical pool to explore cross-cultural comparisons. 

\section{Conclusion}
Within this paper, we present the results of an online survey with 776 people on their everyday use of LLMs for learning. Guided by four research questions, we first examined demographic differences between the 88\% of learners and 11\% of users of LLMs and those 2\% who do not use LLMs at all. We then examined motivations and challenges associated with using LLMs for everyday learning, as well as perceptions of over-reliance, privacy, and productivity among LLM learners. Finally, we identified four types of LLM learners based on the contexts, devices, and tasks involved in using LLMs for everyday learning, namely, \textit{Structured
Knowledge Builders, Self-Guided Explorers, Analytical Problem Solvers}, and \textit{Adaptive Power Users}. Our findings reveal that primarily male adults aged 25–30 are at the forefront of adopting LLMs for self-directed learning. They use LLMs regardless of time and date in a variety of contexts and for a large span of tasks, with the goal to deepen their understanding, explore new subjects, and overcome traditional learning barriers. Our results reveal a paradox: respondents use LLMs for fact-checking while being skeptical about their accuracy -- with a similar narrative for privacy. Moreover, our respondents do not perceive themselves as being overly reliant on LLMs. We conclude with implications that emphasize the importance of including different media types for learning, enabling collaborative learning, providing sources, and catering to specific types of learners and learning \textit{by design.}  

%
%


\bibliographystyle{ACM-Reference-Format}
\bibliography{bibliography}

\appendix

\section{Demographics}

\small
\begin{table}[h]
    \centering
    \caption{Age Distribution}
    \begin{tabular}{lrrrrrr}
        \toprule
        Group & Count & Mean & Std. Dev. & Min & Median & Max \\
        \midrule
        All & 776 & 31.56 & 9.87 & 18 & 29 & 72 \\
        Non-LLM Users & 15 & 35.00 & 11.06 & 22 & 36 & 61 \\
        Non-LLM Learners & 83 & 34.93 & 11.34 & 19 & 33 & 68 \\
        LLM Learners & 678 & 31.07 & 9.57 & 18 & 29 & 72 \\
        \bottomrule
    \end{tabular}
    \label{tab:age-distribution}
\end{table}

\begin{table}[h]
    \centering
    \caption{Gender Breakdown (\%)}
    \begin{tabular}{lccccc}
        \toprule
        Group & Male & Female & Non-binary & Prefer not to say & Self-describe \\
        \midrule
        All & 49.9\% & 48.5\% & 1.0\% & 0.5\% & 0.1\% \\
        Non-LLM Users & 26.7\% & 66.7\% & 6.7\% & 0\% & 0\% \\
        Non-LLM Learners & 30.1\% & 63.9\% & 4.8\% & 0\% & 1.2\% \\
        LLM Learners & 52.8\% & 46.2\% & 0.4\% & 0.6\% & 0\% \\
        \bottomrule
    \end{tabular}
    \label{tab:gender-breakdown}
\end{table}

\begin{table}[h]
    \centering
    \caption{Education Level Distribution (\%)}
    \begin{tabular}{lcccc}
        \toprule
        Group & Bachelors & Graduate Degree & Some University & Completed Secondary \\
        \midrule
        All & 32.2\% & 24.9\% & 17.3\% & 13.4\% \\
        Non-LLM Users & 33.3\% & 6.7\% & 13.3\% & 13.3\% \\
        Non-LLM Learners & 31.3\% & 30.1\% & 16.9\% & 10.8\% \\
        LLM Learners & 32.3\% & 24.6\% & 17.4\% & 13.7\% \\
        \bottomrule
    \end{tabular}
    \label{tab:education-level}
\end{table}

\begin{table}[h]
    \centering
    \caption{Household Income Distribution (\%)}
    \begin{tabular}{lcccccc}
        \toprule
        Group & $<$25K & 25K--50K & 50K--100K & 100K--200K & $>$200K & Prefer not to say \\
        \midrule
        All & 27.4\% & 26.7\% & 29.1\% & 9.2\% & 0.4\% & 7.2\% \\
        Non-LLM Users & 40.0\% & 46.7\% & 6.7\% & 0\% & 0\% & 6.7\% \\
        Non-LLM Learners & 30.1\% & 25.3\% & 27.7\% & 4.8\% & 0\% & 12.0\% \\
        LLM Learners & 26.9\% & 26.4\% & 29.8\% & 9.9\% & 0.4\% & 6.6\% \\
        \bottomrule
    \end{tabular}
    \label{tab:household-income}
\end{table}

\begin{table}[h]
    \centering
    \caption{AI Literacy Scores}
    \begin{tabular}{lrrrrrrr}
        \toprule
        Group & Count & Mean & Std. Dev. & Min & 25\% & Median & Max \\
        \midrule
        All & 776 & 3.06 & 1.25 & 0.00 & 2.00 & 3.00 & 5.00 \\
        Non-LLM Users & 15 & 2.87 & 0.92 & 1.00 & 2.00 & 3.00 & 4.00 \\
        Non-LLM Learners & 83 & 3.22 & 1.36 & 0.00 & 2.00 & 3.00 & 5.00 \\
        LLM Learners & 678 & 3.04 & 1.24 & 0.00 & 2.00 & 3.00 & 5.00 \\
        \bottomrule
    \end{tabular}
    \label{tab:ai-literacy}
\end{table}

\begin{table}[h]
    \centering
    \caption{ATI Scale}
    \begin{tabular}{lrrr}
        \toprule
        Group & Mean & Std. Dev. & Cronbach's Alpha \\
        \midrule
        All & 3.97 & 0.88 & 0.90 \\
        Non-LLM Users & 3.25 & 1.01 & 0.93 \\
        Non-LLM Learners & 3.57 & 0.98 & 0.93 \\
        LLM Learners & 4.04 & 0.84 & 0.90 \\
        \bottomrule
    \end{tabular}
    \label{tab:ati-scale}
\end{table}

\begin{table}[h]
    \centering
    \caption{Personality Traits (Mean Scores)}
    \begin{tabular}{lccccc}
        \toprule
        Group & Extraversion & Agreeableness & Conscientiousness & Neuroticism & Openness \\
        \midrule
        All & 2.85 & 3.65 & 3.42 & 2.92 & 3.60 \\
        Non-LLM Users & 2.50 & 3.29 & 3.57 & 3.53 & 3.57 \\
        Non-LLM Learners & 2.70 & 3.73 & 3.33 & 3.08 & 3.73 \\
        LLM Learners & 2.87 & 3.65 & 3.43 & 2.89 & 3.58 \\
        \bottomrule
    \end{tabular}
    \label{tab:personality-traits}
\end{table}

\section{LCA Model Fit by Number of Classes}

\begin{center}
    \includegraphics[width=.5\linewidth]{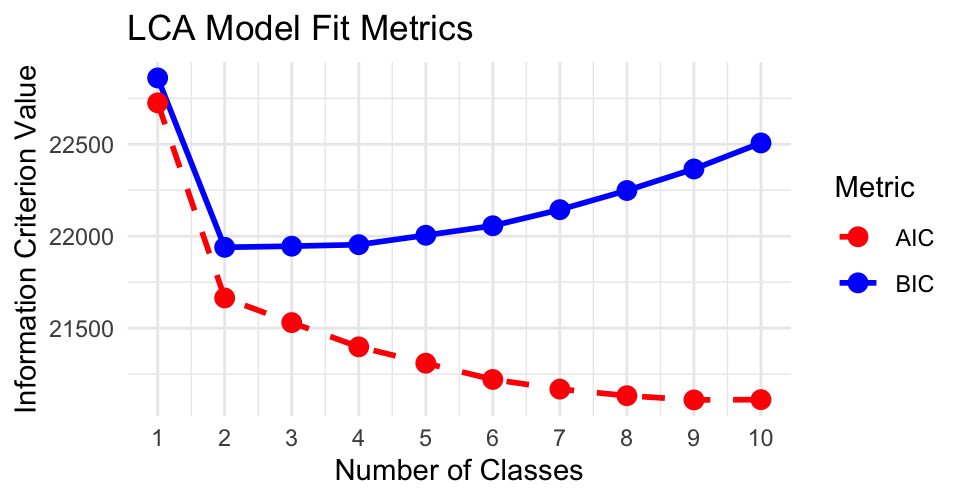}
    \label{fig:lca-fit}
\end{center}

\newpage
\section{Descriptive Statistics of Resulting LCA Classes}

\begin{table}[ht]
  \centering
  \caption{Age Distribution Descriptive Statistics}
  \label{tab:age_stats}
  \begin{tabular}{lrrrr}
    \toprule
    Statistic & Structured Knowledge Builders & Self-Guided Explorers & Analytical Problem Solvers & Adaptive Power Users \\
    \midrule
    Count     & 155.00                         & 180.00                & 167.00                     & 176.00               \\
    Mean      & 28.54                          & 33.99                 & 31.20                      & 30.18                \\
    Median    & 27.00                          & 31.00                 & 29.00                      & 28.00                \\
    Std.\ Dev.& 8.15                           & 10.82                 & 9.48                       & 8.69                 \\
    Min       & 18.00                          & 18.00                 & 18.00                      & 18.00                \\
    Max       & 64.00                          & 72.00                 & 68.00                      & 59.00                \\
    \bottomrule
  \end{tabular}
\end{table}

\begin{table}[ht]
  \centering
  \caption{Gender Distribution Counts by Segment}
  \label{tab:gender_counts}
  \begin{tabular}{lrrrr}
    \toprule
    Gender                     & Structured Knowledge Builders & Self-Guided Explorers & Analytical Problem Solvers & Adaptive Power Users \\
    \midrule
    Female                     & 98                            & 93                    & 43                         & 79                  \\
    Male                       & 56                            & 87                    & 120                        & 95                  \\
    Non-binary  & 0                             & 0                     & 2                          & 1                   \\
    Prefer not to say          & 1                             & 0                     & 2                          & 1                   \\
    \bottomrule
  \end{tabular}
\end{table}

\begin{table}[ht]
  \centering
  \caption{AI Literacy Descriptive Statistics}
  \label{tab:ai_literacy_stats}
  \begin{tabular}{lrrrr}
    \toprule
    Statistic & Structured Knowledge Builders & Self-Guided Explorers & Analytical Problem Solvers & Adaptive Power Users \\
    \midrule
    Count     & 155.00                         & 180.00                & 167.00                     & 176.00               \\
    Mean      & 2.95                           & 2.96                  & 3.17                       & 3.07                 \\
    Std.\ Dev.& 1.26                           & 1.35                  & 1.18                       & 1.15                 \\
    \bottomrule
  \end{tabular}
\end{table}

%
%
\end{document}